\documentclass[12pt,a4paper]{article}
\usepackage{graphicx,times}
\usepackage{amsmath}
\usepackage{amsfonts}
\usepackage{amssymb}
\usepackage{tikz}
\usepackage{pgflibraryarrows}
\usepackage{colortbl}
\usepackage{amsmath}
\usepackage{latexsym}
\usepackage{tikz}
\usepackage{pgflibraryarrows}
\usepackage{tabularx}
\usepackage{appendix}
\usepackage[T1]{fontenc}
\usepackage{longtable}
\usepackage{url}

\begin{document}
\title{Analysing Networks of Networks\thanks{This work was supported by funding from the Russian Science Foundation (project No. 19-18-00394, `Creation of knowledge on ecological hazards in Russian and European local communities'). The work of Jones was carried out while being an academic visitor to The Social Networks Lab, The University of Melbourne}}
\author{Johan Koskinen\thanks{The Social Networks Lab, The Melbourne School of Psychological Sciences, University of Melbourne johan.koskinen@munimelb.edu.au. } \and Pete Jones\thanks{The Mitchell Centre for Social Network Analysis, pete.jones@unimelb.edu.au} \and Darkhan Medeuov\thanks{Centre for German and European Studies, St Petersburg University}, \and Artem Antonyuk\footnotemark[4] \and Kseniia Puzyreva\footnotemark[4] \and Nikita Basov \footnotemark[4] }


\maketitle
\begin{abstract}
We consider data with multiple observations or reports on a network in the case when these networks themselves are connected through some form of network ties. We could take the example of a cognitive social structure where there is another type of tie connecting the actors that provide the reports; or the study of interpersonal spillover effects from one cultural domain to another facilitated by the social ties. Another example is when the individual semantic structures are represented as semantic networks of a group of actors and connected through these actors' social ties to constitute knowledge of a social group. How to jointly represent the two types of networks is not trivial as the layers and not the nodes of the layers of the reported networks are coupled through a network on the reports. We propose to transform the different multiple networks using line graphs, where actors are affiliated with ties represented as nodes, and represent the totality of the different types of ties as a multilevel network. This affords studying the associations between the social network and the reports as well as the alignment of the reports to a criterion graph. We illustrate how the procedure can be applied to studying the social construction of knowledge in local flood management groups. Here we use multilevel exponential random graph models but the representation also lends itself to stochastic actor-oriented models, multilevel blockmodels, and any model capable of handling multilevel networks.
\end{abstract}

\section{Introduction and basic definitions}
The use of graphs in social network research has served to give analytical clarity to a vast range of complex phenomena. Not only have graphs been used to make sense of relationships among the same types of entities in one-mode networks but also between different types of entities in two-mode networks (as early as Hobson 1894/1954). Joint representation of different types of ties among different types of nodes in the combination of one-mode and two-mode networks (Wasserman and Iacobucci, 1991; Snijders, Lomi, and Torl\'{o}, 2013), help us understand how for example social ties determine and are determined by preferences, or affiliations to organisations. Multilevel networks (Lazega et al., 2008) provide a simple and elegant way of representing how interactions among one type of node relate to interactions among other types of nodes. Multiple networks may also be connected pairwise through ties between nodes in different layers or slices (Kivel{\"a} et al., 2014). These simple extensions or refinements of binary simple graphs provide a powerful set of tools for making sense of complex interrelations across multiple settings.

Both multilevel networks and multilayered networks link one-mode networks through ties between the nodes in respective graphs. In a multilevel network, you can link a collaboration network among people and an inter-organisational network by connecting the people with the organisations they belong to. In a multilayered network, if you have networks observed over time, you can connect the pairs of nodes, in consecutive observations, that represent the same individuals. However, if the multiple networks correspond to, say, the semantic networks of different individuals and you want to investigate the extent to which people that are friends tend to connect the same concepts, the relevant connection between networks is not between the nodes. Similarly, if multiple raters provide their versions of a network and we know that the raters themselves are connected in a social network, how do we investigate whether relationally tied raters tend to report ties among the same people? Here, we present a framework for organising the information in a way that provides some analytical clarity and that facilitates modelling of these type of data.
 
We define the network on which we have multiple observations, as graphs on a set of nodes $\mathcal{N}=\{1,\ldots,N \}$ with a set of possible ties $\mathcal{E}=\binom{\mathcal{N} }{2}$. For the multiple reports we assume an index set $V=\{1,\ldots,n\}$, and for each $i \in V$ we assume that we have one report $\mathcal{H}_i=\mathcal{H}_i(\mathcal{N},E_i)$, $E_i \subseteq \mathcal{E}$. This gives us a collection $\mathcal{H}=\{ \mathcal{H}_1,\ldots,\mathcal{H}_n \}$, of networks on the same node set $\mathcal{N}$. We could view $\mathcal{H}$ as a (massively) multiplex network. In addition, we assume that the index set is connected through a network, in which case $V$ becomes nodes that are connected through a network $\mathcal{M}=\mathcal{M}(V,T)$, and where the ties are among the nodes of $V$, $T \subseteq \binom{V }{2}$, and not the nodes in $\mathcal{N}$. We will also allow for the existence of a fixed and given criterion graph $\mathcal{G}( \mathcal{N},E)$ on $\mathcal{N}$. If the $\mathcal{H}_i$'s represent how individuals associate concepts and $\mathcal{M}$ is a social network, there are clearly meaningful ways in which the associations may be informed by social interaction or the other way around, but how to explicate these dependencies in network terms is not straightforward as this is essentially a \emph{network} $\mathcal{M}$, of \emph{networks} $\mathcal{H}_1,\ldots,\mathcal{H}_n$.

We propose to represent the totality of ties as a multilevel network (Lazega et al., 2008) on two node sets, where one node set is $V$ and the other is the set of pairs $\mathcal{E}$ of $\mathcal{N}$ treated as a set of nodes. The latter is key to retaining all information, and by  recognising that $\mathcal{E}$, treated as a set of nodes, means that these nodes are not independently defined but connected through the shared, constituent nodes in $\mathcal{N}$. This is achieved by letting these, latter connections, be represented through a line-graph (Harary and Norman, 1960) $\mathcal{Q}$ on the node set $\mathcal{E}$. The new multilevel network thus retains all the relevant information about the network of networks as well as the potential role of the criterion graph.

There is a long tradition of analysing multiplex networks (Mitchell, 1974; Davis, 1968; Wasserman and Faust, 1994), with many of the classic network datasets (Sampson, 1969; Kapferer, 1972; Padgett and Ansell, 1993) specifically designed to  investigate the interrelation of different types of ties. For a multiplex network, $\mathcal{H}$ represents different relations on $\mathcal{N}$. The networks $\mathcal{H}$ may also represent repeated observations on the same type of tie through time or by different reporters. An example of the latter is when multiple raters report on the same relation, such as the cognitive social structures (CSS) data collection paradigm (Moreno, 1934; Newcomb, 1961; Krackhardt, 1987). The challenges associated with the joint analysis of multiplex networks are well documented (White 1963; White et al., 1976; Pattison 1993; Lazega and Pattison, 1999; Rivero Ostoic, 2020; Krivitsky et al., 2020) as are the issues with analysing multiple raters without having a gold standard network on $\mathcal{N}$ for reference in CSS (see e.g., Butts, 2003). Nevertheless, a wealth of research has come out of the study of multiple networks in the form of $\mathcal{H}$. For multiplex analysis, $n$ is typically small, enabling the researcher to formulate specific hypotheses about how the networks $\mathcal{H}$ are related (Koehly and Pattison, 2005). Large $n$ leads to a combinatorial explosion that makes investigating theoretically informed multiplexities difficult but large $n$ still permits exploration of dimensions of relations (V\"{o}r\"{o}s and Snijders, 2017); relational algebras (Rivero Ostoic, 2017, 2020) and testing of structure using a multigraph representation (Shafie, 2015). For CSS $n=N$ and the main challenge is typically how to relate $\mathcal{H}$ to an unobserved consensus structure or criterion graph $\mathcal{G}(\mathcal{N},E)$ but could we model the dependencies among the $\mathcal{H}_1,\ldots,\mathcal{H}_n$ directly while also accounting for some additional, directly observed relation $\mathcal{M}$ amongst them?

There is a growing number of conceptual frameworks for joint analysis of different types of ties over different types of nodes, such as multilevel networks (Lazega et al., 2008),  multilayered networks (Kivel{\"a} et al., 2014), sociosemantic networks (Basov, 2020), 
socioecological networks (Bodin et al., 2016), etc. Many on them have in common that networks on different types of nodes are connected through two-mode affiliation ties. In particular, a network $\mathcal{G}$ is connected to a network $\mathcal{M}$, through affiliations linking the nodes in $V$ to the nodes in $\mathcal{N}$. This affiliation network consequently connects a social network with another type of network, on a different node set $\mathcal{N}$. If we have multiple networks on $\mathcal{N}$, and that it is these networks that are connected through social ties, the multilevel one-mode by two-mode representation does not apply straightforwardly. For multilevel networks, there are instances where multiple networks on $\mathcal{N}$ may add insight over and above the representation using affiliations of $V$ to nodes in $\mathcal{N}$. For example, Wang et al. (2015), study consumer preferences among products in a product layer, network of similarities among products, and a social network among consumers in a sociomaterial network (Contractor et al., 2011). In the canonical representation of clans, forestry, and ecology as a socioecological network (Bodin and Teng\"{o}, 2012), the clans are assumed to act on a universally understood network $\mathcal{G}$ of forests. This ecological network $\mathcal{G}$ of forests could be disaggregated and represented as each clan's perception or understanding $\mathcal{H}_1,\ldots,\mathcal{H}_n$ of how forests are related. An objectively true network among forests could still be modelled as a criterion graph. In sociosemantic networks (Roth and Cointet, 2010; Hellsten and Leydesdorff, 2017; Basov, Lee, and Antoniuk, 2016; Basov, 2020), the semantic network may be taken as a normative, exogenously given network $\mathcal{G}$ that social actors relate to through affiliations with concepts. If the semantic network is a local semantic network, aggregated across individual meaning structures (the personal `semantic networks') $\mathcal{H}_1,\ldots,\mathcal{H}_n$, this local semantic network could be disaggregated into $n$ versions of how the concepts are related.

Multiplex approaches can handle complex dependencies between and within $\mathcal{H}$ for small $n$. CSS may handle the comparison of $\mathcal{H}$ and a criterion graph $\mathcal{G}$ for large $n$ assuming independence between and within reports $V$, conditional on $\mathcal{G}$. Accounting for complex dependencies within and between $\mathcal{H}$ whilst also comparing these to a criterion graph $\mathcal{G}$, is thus a considerable challenge. In particular, how do we account for dependence between reports, conditionally on $\mathcal{G}$ if these dependencies are induced by a network $\mathcal{M}$ on $V$? Here we propose mapping $\{ \mathcal{H} , \mathcal{G}, \mathcal{M} \}$ to a multilevel network (Lazega et al., 2008) on $V$ and $\mathcal{E}$, with ties defined by $\mathcal{M}$ and $\mathcal{H}$, and $\mathcal{G}$ represented by node-level covariates. We demonstrate how meaningful hypotheses for the original representation of data may be translated into hypotheses expressed in terms of \emph{configurations} (Moreno and Jennings, 1938) in the new, multilevel form. The multilevel representation of the network of networks means that a network of networks lends itself to estimation using any model or statistical package that can model multilevel networks, such as stochastic actor-oriented models (Hollway et al., 2017), exponential random graph models (Wang et al., 2013), and blockmodels (\v{Z}ibena and Lazega, 2016). We illustrate the application of the representation to test hypotheses about the local production of knowledge for a dataset on flood management. We find dyadic social network effects on knowledge: People who are socially connected also tend to connect the same concepts via meaningful associations, i.e., to generate knowledge jointly. Moreover, the multilevel representation of network of networks approach allows us to show  that alignment with a normative semantic network (here, of expert knowledge) moderates local social production of knowledge. This opens new methodological prospects for studying the dispositional effects on local production of culture (Rawlings and Childress, 2019)


\section{Multilevel representation}
When having one-mode networks among a set of actors that are also affiliated with organisations, that, in turn, are connected amongst themselves, Lazega et al. (2008) propose to represent the totality of ties in a multilevel network. A multilevel network thus has two types of nodes and three types of ties. While the multilevel network could be represented as one one-mode network, where the types of ties were identified by their constituent nodes, Lazega et al. (2008) demonstrate the analytical advantages that can be had from retaining a strict distinction between the networks. In particular, the multilevel representation affords specifying different types of dependencies depending on the combination of the types of ties that are used. Table~\ref{tab:notation} in the Appendix provides a list of the notation used in the sequel along with some brief examples of what these might correspond to.

\subsection{Network of networks}
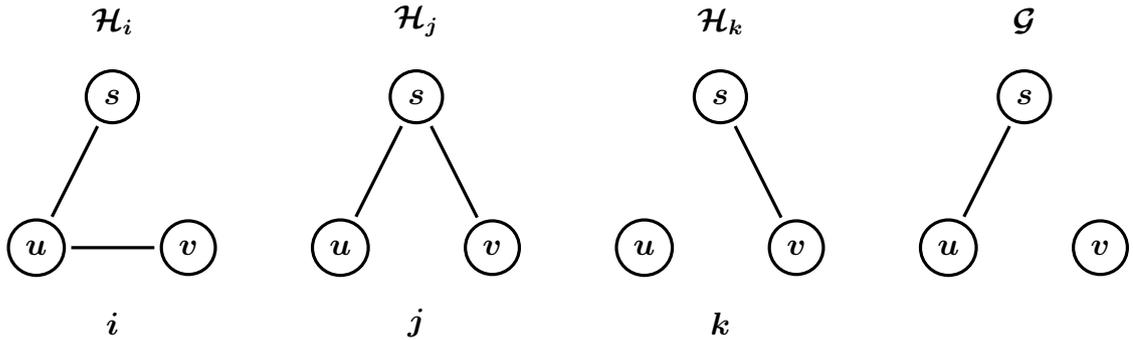
\begin{figure}[ht]
\begin{center}
\begin{tikzpicture}
\begin{scope}

\boldmath
\tikzstyle{every
node}=[shape=circle,minimum size=2ex]
\tikzstyle{every path}=[very thick, -stealth', shorten <=2pt,
shorten >=2pt, -]
\node (Hi) at (1,5)  {{$\mathcal{H}_i$}} ;
\node[draw] (ui) at (0,2)  {{$u$}} ;
\node[draw] (si) at (1,4)  {{$s$}} ;
\node[draw] (vi) at (2,2)  {{$v$}} ;
\node (i) at (1,1)  {{$i$}} ;

\node (Hj) at (5,5)  {{$\mathcal{H}_j$}} ;
\node[draw] (uj) at (4,2)  {{$u$}} ;
\node[draw] (sj) at (5,4)  {{$s$}} ;
\node[draw] (vj) at (6,2)  {{$v$}} ;
\node (j) at (5,1)  {{$j$}} ;

\node (Hk) at (9,5)  {{$\mathcal{H}_k$}} ;
\node[draw] (uk) at (8,2)  {{$u$}} ;
\node[draw] (sk) at (9,4)  {{$s$}} ;
\node[draw] (vk) at (10,2)  {{$v$}} ;
\node (k) at (9,1)  {{$k$}} ;

\node (G) at (13,5)  {{$\mathcal{G}$}} ;
\node[draw] (u) at (12,2)  {{$u$}} ;
\node[draw] (s) at (13,4)  {{$s$}} ;
\node[draw] (v) at (14,2)  {{$v$}} ;


\draw[very thick]  (ui) -- (si);  \draw[very thick]  (ui) -- (vi);  
\draw[very thick]  (uj) -- (sj);  \draw[very thick]  (sj) -- (vj);  
 \draw[very thick]  (sk) -- (vk);  
  \draw[very thick]  (u) -- (s);  

\end{scope}
\end{tikzpicture}

\end{center}
\caption{ Networks of $i,j,k \in V$ and criterion network $\mathcal{G}$}
\label{fig:css}
\end{figure}

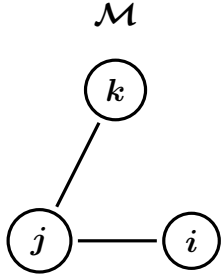
\begin{figure}[ht]
\begin{center}
\begin{tikzpicture}
\begin{scope}

\boldmath
\tikzstyle{every
node}=[shape=circle,minimum size=2ex]
\tikzstyle{every path}=[very thick, -stealth', shorten <=2pt,
shorten >=2pt, -]
\node (Hi) at (1,5)  {{$\mathcal{M}$}} ;
\node[draw] (ui) at (0,2)  {{$j$}} ;
\node[draw] (si) at (1,4)  {{$k$}} ;
\node[draw] (vi) at (2,2)  {{$i$}} ;
\draw[very thick]  (ui) -- (si);  \draw[very thick]  (ui) -- (vi);  

\end{scope}
\end{tikzpicture}

\end{center}
\caption{ A social network amongst nodes $i$, $j$, and $k$ in $\mathcal{M}$}
\label{fig:socnet}
\end{figure}

In the original representation of data used here $\{ \mathcal{H} , \mathcal{G}, \mathcal{M} \}$, there is a clear distinction between the nodes on $\mathcal{G}$ and the $\mathcal{H}_i$'s, on the one hand, and the nodes of $\mathcal{M}$ on the other. The nodes of $V$ have ties amongst themselves but they do not have ties to the nodes of $\mathcal{G}$, something which makes the joint representation less than trivial. To define a joint representation, consider the representations of the original data $\mathcal{H}$ and $\mathcal{G}$ in Figure~\ref{fig:css} and $\mathcal{M}$ in Figure~\ref{fig:socnet}. Again, the networks $\mathcal{H}_1,\ldots,\mathcal{H}_n$, may be the cognitive representation of informants $i\in V$ of a network on $\mathcal{N}$, or they could be actual observed networks across multiple settings or contexts. These networks can be represented by their adjacency matrices. If each report $\mathcal{H}_i$ is represented by the $N\times N$ adjacency matrix $\mathbf{X}_i$, that has element $X_{iuv}$ equal to 1 if $\{u,v\}\in \mathcal{H}_i$, and 0 otherwise, we can represent $\mathcal{H}$ by the three-way array $\mathbf{X}=\langle\mathbf{X}_i\rangle$ as in Batchelder et al. (1997). The criterion graph similarly is represented by the binary $N\times N$ matrix $\mathbf{A}$. Finally, the social network $\mathcal{M}$ will be denoted by the $n \times n$ binary adjacency matrix $\mathbf{Y}$, where $Y_{ij}$ equal to 1 is $\{i,j\}\in \mathcal{M}$, and 0 otherwise.


The networks in Figure~\ref{fig:css} could represent a multiplex network where $i$, $j$, and $k$, are different relations and $\mathcal{G}$ an additional fourth relation. In the CSS framework the nodeset $\mathcal{N}$ would be the same as $V$, and $\mathcal{G}$ would represent some consensus graph on the same node set. In the case of multiplex networks and $n$ small, we could formulate specific hypotheses for how the networks $\mathcal{H}$ relate pair-wise to each other. For example in the case of generalised exchange, we might ask if a tie $\{s,v\}$ in $\mathcal{H}_k$ closes an open triad $\{\{s,u\},\{u,v\} \}$ in $\mathcal{H}_i$. For CSS, we may introduce dependencies on $V$ through, for example, asserting that respondents are more accurate when reporting on their own ties (Krackhardt, 1987; Batchelder et al. 1997; Butts, 2003; Koskinen, 2004). If $i=u$ and $j=s$, then affording greater accuracy for self-reports would mean that we would trust the reports by $i$ and $j$ on the tie $\{u,s\}=\{i,j\}$, that the tie is present, than the report of $k$, $k\neq i,j$, that reports the tie as being absent.

For networks where $\mathcal{N}$ is not a set of social actors, there are several examples of multiple, `parallel' networks. In the framework of Friedkin et al. (2016), $i$, $j$, and $k$ may represent different systems of belief, where beliefs are represented as connections of concepts in $\mathcal{N}$. Friedkin et al. (2016) do no model $\mathcal{H}$ but take these as a small collection of fixed and known belief systems and assume that people may be influenced to change from endorsing one belief system $\mathcal{H}_i$ to another $\mathcal{H}_j$. A related example is when the networks $\mathcal{H}$ are semantic networks representing the local meaning structures (Basov, de Nooy, and Nenko, 2019) of respondents in $V$.

\begin{figure}[ht]
\begin{center}
\begin{tikzpicture}
\begin{scope}

\boldmath
\tikzstyle{every
node}=[shape=circle,minimum size=2ex]
\tikzstyle{every path}=[very thick, -stealth', shorten <=2pt,
shorten >=2pt, -]
\node (Hi) at (1,1)  {{$\mathcal{H}_i$}} ;
\node[draw] (ui) at (0,2)  {{$u$}} ;
\node[draw] (si) at (1,4)  {{$s$}} ;
\node[draw] (vi) at (2,2)  {{$v$}} ;
\node (i) at (2.5,3.5) {};
\node[draw, minimum size=20ex] (iNet) at (1,2.5) {};
\node (yij) at (2,4.75) {{$y_{ij}=1$}};

\node (Hj) at (5,9)  {{$\mathcal{H}_j$}} ;
\node[draw] (uj) at (4,6)  {{$u$}} ;
\node[draw] (sj) at (5,8)  {{$s$}} ;
\node[draw] (vj) at (6,6)  {{$v$}} ;
\node (j) at (4.5,5) {};
\node (j2) at (6,5) {};
\node[draw, minimum size=20ex] (jNet) at (5,6.6) {};
\node (yjk) at (7.5,5) {{$y_{jk}=1$}};
\node (yik) at (5,1) {{$y_{ik}=0$}};

\node (Hk) at (9,1)  {{$\mathcal{H}_k$}} ;
\node[draw] (uk) at (8,2)  {{$u$}} ;
\node[draw] (sk) at (9,4)  {{$s$}} ;
\node[draw] (vk) at (10,2)  {{$v$}} ;
\node (k) at (7.7,3.5) {};
\node[draw, minimum size=20ex] (kNet) at (9,2.5) {};


\draw[very thick]  (ui) -- (si);  \draw[very thick]  (ui) -- (vi);  
\draw[very thick]  (uj) -- (sj);  \draw[very thick]  (sj) -- (vj);  
 \draw[very thick]  (sk) -- (vk); 
  \draw[very thick]  (i) -- (j);  \draw[very thick]  (j2) -- (k);


\end{scope}
\end{tikzpicture}

\end{center}
\caption{ A Network of networks where $\mathcal{M}$, with adjacency matrix $\mathbf{Y}$, connects the networks on $\mathcal{N}$ of $i,j,k \in V$}
\label{fig:netofnets}
\end{figure}
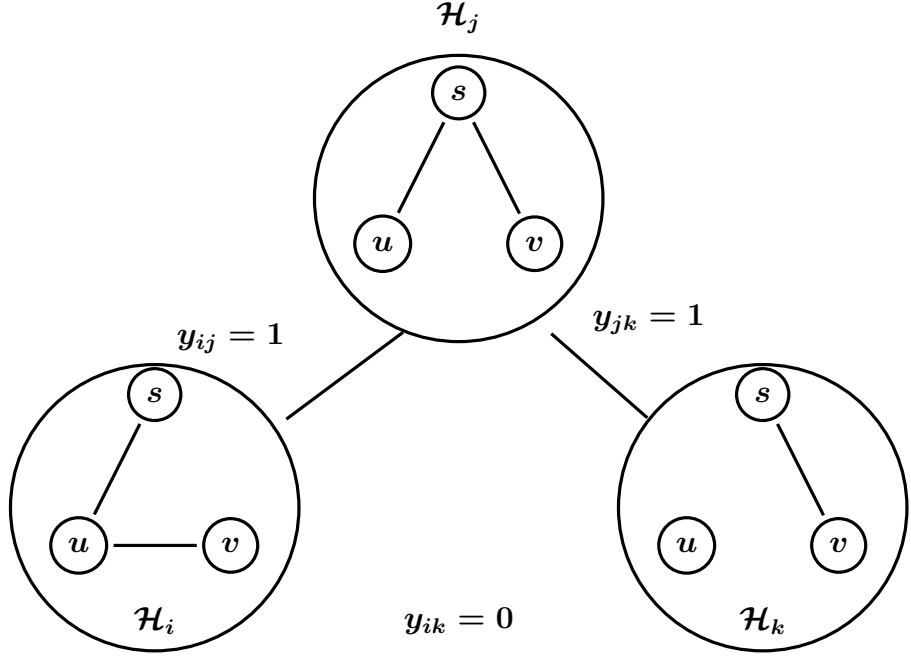

Assuming a network $\mathcal{M}$ as in Figure~\ref{fig:socnet}, means that we have a network amongst the elements of the index set $V$ of $\mathcal{H}$. If $V=\mathcal{N}$, that is, we are only dealing with networks on one type of node, we could represent $\mathcal{M}$ by $\mathcal{G}$ but this does not help us specify how the slices in $\mathcal{H}$ depend on each other. There is nothing in $\mathcal{G}$ in Figure~\ref{fig:css} that, for example, connects, say, $\mathcal{H}_i$ and $\mathcal{H}_j$. The relations between the $\mathcal{H}_i$'s as prescribed by $\mathcal{M}$ is in fact a network of networks (NoNs), as represented in Figure~\ref{fig:netofnets}, where the variables $y_{ij}$, $y_{ik}$, and $y_{jk}$ represent the tie-variables of the adjacency matrix $\mathbf{Y}$ of $\mathcal{M}$. This superficially resembles the representation by Friedkin et al. (2016), that models $n$ individuals' $m$ ($n \neq m$) truth statements (a), as a function of a social network (b) among the individuals and a number of belief systems (c). The social network (b) is a weighted version of $\mathcal{M}$ but here we do not have a structure (a) connecting individuals to belief systems. The belief systems (c) are only superficially similar to the $\mathcal{H}_i$'s and are taken to be exogenously defined and not indexed by the actors. The Friedkin et al. (2016) model is not designed to model the network of networks as we conceive of it.

Further examples may include the individual product preferences of consumers linked through social ties (Wang et al., 2015), perceived food webs by fishers (Barnes et al., 2019), intersectional flows in different countries (Leoncini at al., 1996), symptom networks (Borsboom and Cramer, 2013) within individuals, etc, all of which may be connected by a network: consumers through social ties, fishers through communications, countries through trade ties, people through social ties.

\subsection{Direct modelling of reports, criterion graph, and social network}
To motivate the proposed representation of the network of networks as a multilevel graph we briefly consider the challenges associated with modelling $\mathcal{H}$ directly while incorporating the possible dependence through $\mathcal{M}$.

As observed above, we may model $\mathcal{H}$ directly as a multiplex network, conditional or unconditional on $\mathcal{G}$. For example, we can assume that $\mathbf{X}_1,\ldots,\mathbf{X}_n$ follow a multiplex exponential random graph model (Lazega and Pattison, 1999) conditional on the adjacency matrix $\mathbf{A}$ of $\mathcal{G}$, with parameters $\theta$
\[
p_{\theta}(\mathbf{X}_1,\ldots,\mathbf{X}_n| \mathbf{A} )=\exp\{ \theta^{\top} z(\mathbf{X}_1,\ldots,\mathbf{X}_n; \mathbf{A} ) - \psi(\theta) \}
\]
for a normalising constant $\psi(\theta)$, and where the vector of statistics $z(\cdot)$ has components
\[
z_{k_1,\ldots,k_r}( z(\mathbf{X}_1,\ldots,\mathbf{X}_n; \mathbf{A} ) )=z_{k_1,\ldots,k_r}( z(\mathbf{X}_{k_1},\ldots,\mathbf{X}_{k_r}; \mathbf{A} ) )
\]
that are functions of subsets
\[
\{ k_1,\ldots,k_r \} \in \binom{ V }{r}
\]
For $r=2$ we can specify statistics corresponding to entrainment and generalised exchange through terms of the type $\sum_{k,h} X_{ikh}X_{jkh}$ and $\sum_{k,h,\ell}X_{ikh}X_{ih\ell}X_{jk\ell}$. Dependence on $\mathcal{M}$ can be introduced through interactions with the variables of $\mathbf{Y}$. For example, assume that there is alignment between ties in $\mathcal{H}_i$ and $\mathcal{H}_j$ only if $Y_{ij}=1$, in which case the entrainment terms in the multiplex ERGM would be $y_{ij}\sum_{k,h} X_{ikh}X_{jkh}$. Note that when $r=2$, $Y_{ij}$ is either one or zero, and the interaction  $y_{ij}\sum_{k,h} X_{ikh}X_{jkh}$ is either zero or equal to $\sum_{k,h} X_{ikh}X_{jkh}$.   The number of possible statistics grows very quickly with $r$ and the types of statistics we can specify are limited in terms of the types of dependencies we may specify.  Additionally, the direct multiplex ERGM does not afford simultaneous modelling of $\mathcal{H}$ and $\mathcal{M}$, at least not easily or in a practical way.

Statistical models for CSS (Batchelder et al., 1997; Butts, 2003; Koskinen, 2004) have typically assumed that the variables $X_{ikh}$ and $X_{juv}$ are conditionally independent across $V$ and $\mathcal{E}$, conditional on respondent factors and the criterion graph $\mathcal{G}$. Without loss of generality we can assume that
\[
p(\mathbf{X}| g, \mathbf{A}) = \prod_{kh}\prod_{i}\Pr(X_{ikh}=x_{ikh} |A_{kh},g_i(A_{kh}))
\]
where the $g_i(\cdot)$'s may incorporate different actor accuracies. This model is eminently tractable and we cannot relax the independence assumption, introducing dependence among reports through $\mathbf{Y}$, without losing tractability. Let us consider what happens when we introduce dependence through the ties of  $\mathbf{Y}$. For example, we may want to allow $X_{ikh}$ to depend on $X_{jkh}$ if $Y_{i j}=1$. In a modified model
\[
p(\mathbf{X}| g, \mathbf{A}) = \prod_{kh} \Pr(\mathbf{X}_{\cdot kh}=(X_{ihk})_{i \in V} |A_{kh},g_i(A_{kh}))
\]
this can be accommodated, for example, through assuming that conditionally
\[
\textrm{logit} \Pr(X_{ikh}=1| (X_{ihk})_{j \neq i}, A_{kh}=1 ) = \theta_0+\theta_1 \sum_{j \neq i}Y_{ij}X_{jkh} \text{,}
\] 
an expression that we recognise as the conditional form of the auto-logistic actor attribute model (ALAAM) (Daraganova and Robins, 2013; Koskinen and Daraganova, 2020). In other words, for each $\{ i,j \} \in \mathcal{E}$ we would end up with an ALAAM for the vector of responses $X_{1kk},\ldots,X_{nkh}$, in total $e$ different ALAAMs. ALAAMs might be challenging and would restrict the nature of dependencies we can consider. This framework would for example not allow us to consider dependencies within respondents such as how the responses $X_{ikh}$ and $X_{iuv}$ may or may not be dependent.

\subsection{Representing a network of networks as a multilevel network}
We are able to encode the information in $\mathcal{M}$ and $\mathcal{H}$, and $\mathcal{G}$ in one multilevel network, by representing the dyads $\mathcal{E}$ of $\mathcal{H}$ as nodes. Define a mapping from the set of dyads $\mathcal{E}$ to  $ \mathcal{P}=\{1,\ldots,e \}$, $\pi:\mathcal{E} \rightarrow \mathcal{P}$. For nodes $i \in V$, we define an affiliation matrix on $ V \times \mathcal{P}$ as an $n \times e $ affiliation matrix $\mathbf{W}$ with elements
\begin{equation*}  
	W_{i r } = \left\{
	\begin{array}{lr}
		1,&\text{if } X_{i \pi^{-1}(r) }=1 \\
		0,&\text{otherwise }\\
	\end{array} 
	\right. 
\end{equation*}
Thus, if $i \in V$ has a tie $\{ u,v \} \in \mathcal{H}_i $, and $\pi( \{ u,v \} )= r$, then there is an affiliation tie $W_{ir}=1$. To account for the criterion graph $\mathcal{G}$, the vertices $\mathcal{P}$ are coloured according to whether $\{ u,v \} \in \mathcal{G}$ or not, for $r$ such that $\pi( \{ u,v \} )= r$. The binary colouring of $\mathcal{P}$ is a vector
\begin{equation*}  
	D_{ r } = \left\{
	\begin{array}{lr}
		1,&\text{if } A_{\pi^{-1}(r) }=1 \\
		0,&\text{otherwise }\\
	\end{array} 
	\right. 
\end{equation*}
Denoting the social network on $V$ by $\mathbf{Y}$, as before, and the networks $\mathbf{X}_1,\ldots,\mathbf{X}_n$  expressed as $\mathbf{W}$, we can define a multilevel network $\mathcal{C}$ as a blocked adjacency matrix
\begin{equation*}  
	\underset{(e+n ) \times (e+n )}{ \mathbf{C}} = \left(
	\begin{array}{cc}
		\mathbf{0}_{e \times e  } & \mathbf{W}^{\top}\\
		\mathbf{W} & \mathbf{Y}
	\end{array} 
	\right) 
\end{equation*}
where $\mathbf{0}_{e \times e } $ is a matrix of zeros. The mapping $\pi$ is arbitrary at this point which means that the structure of the network $\mathcal{G}$ is only reflected through $\mathbf{D}$. The representation $\mathbf{C}$ is thus agnostic to whether, for example, $r,t \in \mathcal{P}$ refer to pairs $\pi^{-1}(r)$ and $\pi^{-1}(t)$ that may share a node or not. To relax this independence, we introduce a top-level network by connecting ties that share nodes. Formally, define a graph $\mathcal{Q}$ as the $e \times e$ matrix $\mathbf{Q}$ with elements
\begin{equation*}  
	Q_{ rt } = \left\{
	\begin{array}{lr}
		1,&\text{if } \pi^{-1}(r) \cap  \pi^{-1}(t) \neq \emptyset\\
		0,&\text{otherwise }\\
	\end{array} 
	\right. 
\end{equation*}
The graph $\mathcal{Q}$ has many interpretations. If all pairs of $\mathcal{E}$ are mapped to $\mathcal{P}$, then $\mathcal{Q}$ is the line graph of the complete graph on $\mathcal{N}$. If the mapping $\pi$ is applied only to a subset $E \subset \mathcal{E}$, then $\mathcal{Q}$ is the line graph of a graph on $\mathcal{N}$ with edge set $E$. For any graph on $\mathcal{N}$, $\mathcal{Q}$ is formally the dependence graph on $\mathcal{E}$ under the Markov dependence assumption  (Frank and Strauss, 1987). Additionally, the graph $\mathcal{Q}$  is the complement of the Kneser graph $KG_{N,2}$.\footnote{A Kneser (1955) graph $KG_{n,k}$ has nodes $V= \binom{n}{k}$, and edge set $E= \{ (u,v) \in V: u \cap v = \emptyset \}$. The complement of $KG_{n,s}$, is a line-graph where the edge set is $\bar{E}=\{ (u,v) \in V : u \cap v \neq \emptyset \}$. The graph on $V$ and $\bar{E}$ is exactly the dependence graph, $D$, of a Markov random graph (Frank and Strauss, 1986; Lusher et al., 2013) on $V=\{ 1,\ldots,n \}$.}%

Taken together, we have the blocked adjacency matrix of a multilevel network 
\begin{equation*}  
	\mathbf{C} = \left(
	\begin{array}{cc}
		\mathbf{Q} & \mathbf{W}^{\top}\\
		\mathbf{W} & \mathbf{Y}
	\end{array} 
	\right) 
\end{equation*}

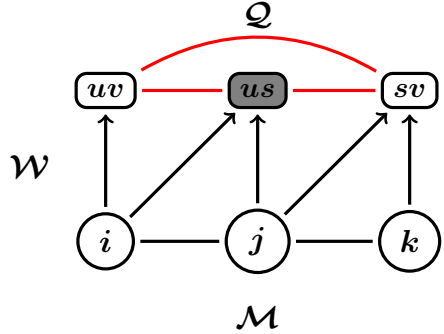
\begin{figure}[ht]
\begin{center}
\begin{tikzpicture}
\begin{scope}
\boldmath
\tikzstyle{every
node}=[shape=circle,minimum size=2ex]
\tikzstyle{every path}=[very thick, -stealth', shorten <=2pt,
shorten >=2pt, -]
\node (W) at (-1,3)  {{$\mathcal{W}$}} ;
\node (M) at (2,1)  {{$\mathcal{M}$}} ;
\node (Q) at (2,5)  {{$\mathcal{Q}$}} ;

\node[draw] (i) at (0,2)  {{$i$}} ;
\node[draw] (j) at (2,2)  {{$j$}} ;
\node[draw] (k) at (4,2)  {{$k$}} ;
\node[draw,shape=rectangle,rounded corners]  (uv) at (0,4) {{\small $uv$}};
\node[draw,shape=rectangle,rounded corners,fill=gray]  (us) at (2,4) {{\small $us$}};
\node[draw,shape=rectangle,rounded corners]  (sv) at (4,4) {{\small $sv$}};
\draw[draw,very thick, -stealth', shorten <=2pt, shorten >=2pt, ->]  (i) -> (uv);   
\draw[draw,very thick, -stealth', shorten <=2pt, shorten >=2pt, ->]  (i) -> (us);  
\draw[draw,very thick, -stealth', shorten <=2pt, shorten >=2pt, ->]  (j) -> (us); 
\draw[draw,very thick, -stealth', shorten <=2pt, shorten >=2pt, ->]  (j) -> (sv);    
\draw[draw,very thick, -stealth', shorten <=2pt, shorten >=2pt, ->]  (k) -> (sv); 

\draw[red, very thick]  (uv) -- (us);  \draw[red, very thick]  (sv) -- (us);  
\draw[red, very thick]  (uv) to [bend left] (sv);   
\draw[very thick]  (i) -- (j);  \draw[very thick]  (j) -- (k);  


\end{scope}
\end{tikzpicture}

\end{center}
\caption{ The multilevel network representation $\mathcal{C}$, of the network of networks in Figure~\ref{fig:netofnets}, consisting of the networks in Figure~\ref{fig:css} connected through the network in Figure~\ref{fig:socnet}. Social actors $i$, $j$, and $k$ are socially connected in $\mathcal{M}$. An actor $i$, has a tie in $\mathcal{W}$ to the links $\{u,v\}$ of their reports $\mathcal{H}_i$. Nodes in $\mathcal{Q}$ are connected if they share a node in $\mathcal{N}$ in the preimage. Nodes in $\mathcal{Q}$ have a binary attribute indicating if they are in the criterion graph or not.}
\label{CovERGM}
\end{figure}
When analysing $\mathcal{C}$ we need to respect the fact that $\mathcal{Q}$ is a fixed and exogenous graph that is completely determined by the index set.

We now proceed to describe how network configurations (Moreno and Jennings, 1938) in $\mathcal{C}$ correspond to meaningful combinations of ties in the network of networks.

\subsubsection{Basic configurations for $\mathcal{H}$ \label{sec:socialprod}}
\begin{figure}[ht]
\begin{center}
\begin{tikzpicture}
\begin{scope}

\boldmath
\tikzstyle{every
node}=[shape=circle,minimum size=2ex]
\tikzstyle{every path}=[very thick, -stealth', shorten <=2pt,
shorten >=2pt, -]
\node[draw] (u) at (0,2)  {{$u$}} ;
\node[draw] (s) at (0,4)  {{$s$}} ;

\node (Hi) at (0,5)  {{$\mathcal{H}_i$}} ;

\node (ilab) at (0,0) {$i$};

\node[draw] (i) at (4,2)  {{$i$}} ;
\node[draw,shape=rectangle,rounded corners]  (us) at (4,4) {{ $us$}};
  \node (C) at (4,5)  {{$\mathcal{C}$}} ;

\draw[draw,very thick]  (u) -- (s); 
\draw[very thick, -stealth', shorten <=2pt, shorten >=2pt, ->]  (i) -> (us);

\end{scope}
\end{tikzpicture}

\end{center}
\caption{ Mapping density in $\mathcal{H}_i$ to $\mathcal{C}$. Counts of ties in the two-mode graph $\mathcal{W}$, correspond to the total number of ties in $\mathcal{H}$. }
\label{fig:activityactor}
\end{figure}
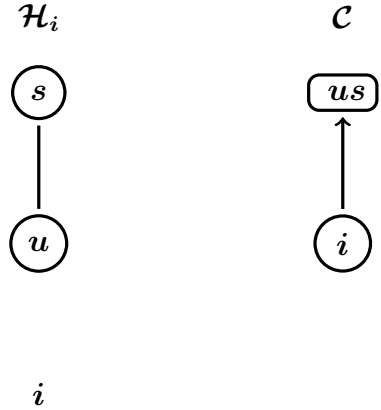

To capture the overall number of ties reported across $i \in V$ corresponds to the  bipartite density as in Figure~\ref{fig:activityactor}. The corresponding count, or statistic is simply $\sum_{i,r}W_{ir}$. Centralisation of ties in $\mathcal{E}$ can be further modelled using bipartite 2-stars $\sum_{r} \binom{W_{+r}}{2}$ or the equivalent alternating stars (Wang et al., 2009). Similarly we may define 2-stars and alternating stars centred on nodes in $V$ (Wang et al., 2009). The majority of configurations we discuss in the sequel have been defined or explored for either two-mode networks (Wang et al., 2009) or multilevel networks (Wang et al., 2013) and we refer the reader to the literature for mathematical definitions of these configurations in $\mathcal{C}$ and focus here on their interpretation in their respective pre-image $\mathcal{H}$, $\mathcal{M}$, and $\mathcal{G}$.

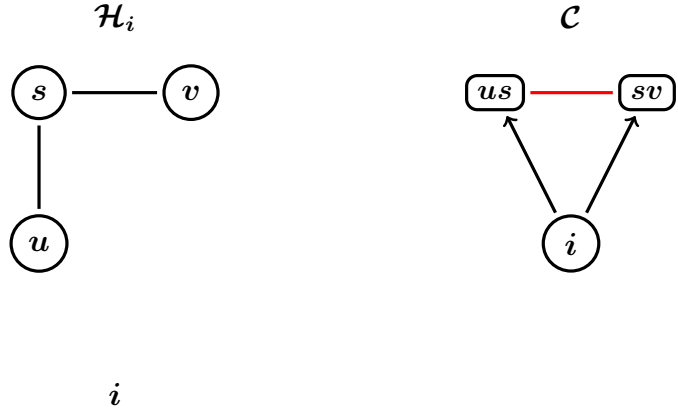
\begin{figure}[ht]
\begin{center}
\begin{tikzpicture}
\begin{scope}

\boldmath
\tikzstyle{every
node}=[shape=circle,minimum size=2ex]
\tikzstyle{every path}=[very thick, -stealth', shorten <=2pt,
shorten >=2pt, -]
\node[draw] (u) at (0,2)  {{$u$}} ;
\node[draw] (s) at (0,4)  {{$s$}} ;
\node[draw] (v) at (2,4)  {{$v$}} ;
\node[draw] (i) at (7,2)  {{$i$}} ;

\node (ilab) at (1,0) {$i$};
\node (Hi) at (1,5)  {{$\mathcal{H}_i$}} ;

\node[draw,shape=rectangle,rounded corners]  (us) at (6,4) {{\small $us$}};
\node[draw,shape=rectangle,rounded corners]  (sv) at (8,4) {{\small $sv$}};
\node (C) at (7,5)  {{$\mathcal{C}$}} ;
\draw[draw,very thick]  (u) -- (s);  \draw[very thick]  (s) -- (v);  

\draw[very thick, -stealth', shorten <=2pt, shorten >=2pt, ->]  (i) -> (us);  
\draw[very thick, -stealth', shorten <=2pt, shorten >=2pt, ->]  (i) -> (sv);

 \draw[red, very thick]  (sv) -- (us);


\end{scope}
\end{tikzpicture}

\end{center}
\caption{ Mapping two-paths in $\mathcal{H}_i$ to multilevel closure in $\mathcal{C}$. For example, centrality of concepts in $\mathcal{H}$ manifests as co-nomination of concept-concept pairs in $\mathcal{Q}$ that share a concept and thus are edges of $\mathcal{Q}$; or, centrality of a person in $\mathcal{H}$ manifests as co-nomination of nodes in $\mathcal{Q}$ that are incident to that person and therefore have a tie in $\mathcal{Q}$.}
\label{fig:spillover}
\end{figure}

Centrality of nodes in $\mathcal{N}$ is reflected in multilevel triangles in $\mathcal{C}$ as in Figure~\ref{fig:spillover}. If the network on $\mathcal{N}$ is a social network, this centralisation reflect the typical heterogeneities that we encounter in social networks, such as preferential attachment. If the nodes on $\mathcal{N}$ are concepts, multilevel closure in $\mathcal{C}$ of the type Figure~\ref{fig:spillover}, could reflect differences in saliency or popularity of concepts but also spill-over (Maki et al., 2019). If individuals associate an energy efficient fridge ($s$) with an energy efficient washing machine ($u$), they may also associate the energy efficient fridge with electric cars ($v$).

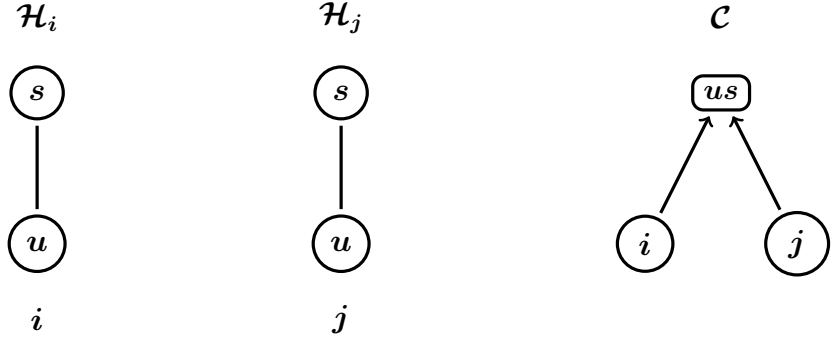
\begin{figure}[ht]
\begin{center}
\begin{tikzpicture}
\begin{scope}

\boldmath
\tikzstyle{every
node}=[shape=circle,minimum size=2ex]
\tikzstyle{every path}=[very thick, -stealth', shorten <=2pt,
shorten >=2pt, -]
\node (Hi) at (0,5)  {{$\mathcal{H}_i$}} ;
\node[draw] (ui) at (0,2)  {{$u$}} ;
\node[draw] (si) at (0,4)  {{$s$}} ;
\node (i) at (0,1)  {{$i$}} ;

\node (Hj) at (4,5)  {{$\mathcal{H}_j$}} ;
\node[draw] (uj) at (4,2)  {{$u$}} ;
\node[draw] (sj) at (4,4)  {{$s$}} ;

\node (j) at (4,1)  {{$j$}} ;

\draw[very thick]  (ui) -- (si); 
\draw[very thick]  (uj) -- (sj);

\node[draw] (i) at (8,2)  {{$i$}} ;
\node[draw] (j) at (10,2)  {{$j$}} ;

\node[shape=rectangle,rounded corners,draw]  (us) at (9,4) {{\small $us$}};
  \node (C) at (9,5)  {{$\mathcal{C}$}} ;
\draw[very thick, -stealth', shorten <=2pt, shorten >=2pt, ->]  (i) -> (us);  
\draw[very thick, -stealth', shorten <=2pt, shorten >=2pt, ->]  (j) -> (us); 

\end{scope}
\end{tikzpicture}

\end{center}
\caption{ Entrainment of ties of $\mathcal{H}_i$ and $\mathcal{H}_j$ expressed as multilevel agreement in $\mathcal{C}$. For example, agreement on the tie $\{u,s\}$ may reflect a cultural consensus over and above any social dependence through $\mathcal{M}$.}
\label{fig:basicentrain}
\end{figure}

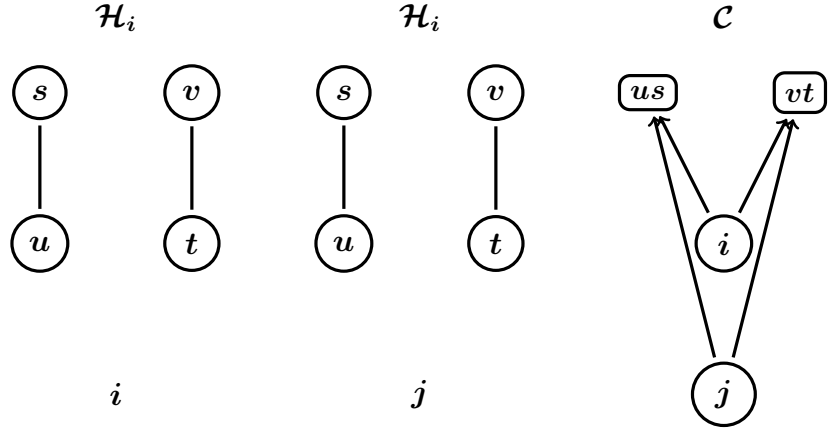
\begin{figure}[ht]
\begin{center}
\begin{tikzpicture}
\begin{scope}

\boldmath
\tikzstyle{every
node}=[shape=circle,minimum size=2ex]
\tikzstyle{every path}=[very thick, -stealth', shorten <=2pt,
shorten >=2pt, -]
\node[draw] (u) at (0,2)  {{$u$}} ;
\node[draw] (s) at (0,4)  {{$s$}} ;
\node[draw] (v) at (2,4)  {{$v$}} ;
\node[draw] (t) at (2,2)  {{$t$}} ;
\node (ilab) at (1,0) {$i$};
\node (Hi) at (1,5)  {{$\mathcal{H}_i$}} ;
\draw[draw,very thick]  (u) -- (s);  \draw[very thick]  (v) -- (t);  

\node[draw] (uj) at (4,2)  {{$u$}} ;
\node[draw] (sj) at (4,4)  {{$s$}} ;
\node[draw] (vj) at (6,4)  {{$v$}} ;
\node[draw] (tj) at (6,2)  {{$t$}} ;
\node (jlab) at (5,0) {$j$};
\node (Hj) at (5,5)  {{$\mathcal{H}_i$}} ;
\draw[draw,very thick]  (uj) -- (sj);  \draw[very thick]  (vj) -- (tj);

\node[draw,shape=rectangle,rounded corners]  (us) at (8,4) {{\small $us$}};
\node[draw,shape=rectangle,rounded corners]  (sv) at (10,4) {{\small $vt$}};
  \node (C) at (9,5)  {{$\mathcal{C}$}} ;
\node[draw] (i) at (9,2)  {{$i$}} ;
\node[draw] (j) at (9,0)  {{$j$}} ;
\draw[very thick, -stealth', shorten <=2pt, shorten >=2pt, ->]  (i) -> (us);  
\draw[very thick, -stealth', shorten <=2pt, shorten >=2pt, ->]  (i) -> (sv); 
\draw[very thick, -stealth', shorten <=2pt, shorten >=2pt, ->]  (j) -> (us);  
\draw[very thick, -stealth', shorten <=2pt, shorten >=2pt, ->]  (j) -> (sv);


\end{scope}
\end{tikzpicture}

\end{center}
\caption{ Association of ties in $\mathcal{E}$ that are not related through the node set $\mathcal{N}$. For example, a consensus that is driven by an endogenous process towards structural equivalence whereby co-nomination by many of one tie leads to agreement on other ties.}
\label{fig:relatedness}
\end{figure}

We could consider a number of ways in which the reports in $\mathcal{H}$ align with each other. A basic form of entrainment is the agreement on a tie between $i$ and $j$ depicted in Figure~\ref{fig:basicentrain}. In a CSS framework we would think of this as relating to the strength of consensus. More generally, this reflects the cultural consensus (Romney et al., 1986; Batchelder and Romney, 1988). While this basic form of entrainment reflects agreement on individual ties in $\mathcal{E}$, co-nomination of pairs of ties in $\mathcal{E}$, as depicted in Figure~\ref{fig:relatedness}, reflects a more structural consensus. The four-cycle in Figure~\ref{fig:relatedness} represents the most basic form of clustering in a two-mode network (Borgatti and Everett, 1997) and is often taken to represent social processes above simple agreement (Robins and Alexander, 2004; Koskinen and Edling, 2012). Here, in very general terms, the configuration can directly be interpreted as whenever two people $i$ and $j$ agree on one thing, they tend to agree on another.

\subsubsection{Social dependencies in $\mathcal{H}$}
Introducing configurations that include the network $\mathcal{M}$ enable us to investigate social construction and social influence as well as homophily induced by shared beliefs. The agreement in Figure~\ref{fig:basicentrain} may be the result of social connections, in which case we expect to see high incidence of the social entrainment configuration of Figure~\ref{fig:socentrain}. For cross-sectional data we cannot tell whether agreement in $\mathcal{W}$ was the result of a social tie in $\mathcal{M}$ or the other way around. The nature of the process explaining configurations as in Figure~\ref{fig:socentrain} is context dependent and may reflect a multitude of processes, such as learning, information, influence, etc. Heider's (1958) \emph{balance theory} is commonly applied in the networks literature in the triadic form of Cartwright and Harary (1956), where two people with a positive tie are assumed likely to also have a positive tie to the same other. The social alignment of Figure~\ref{fig:socentrain}, is a direct application of Heider's (1958) $POX$ scheme, where person $P$ (here $i$) seeks to have ties to other $O$ (here $j$) that like the same object $X$ (here $us$).

Moving beyond direct alignment, we may consider the interactions of multiple types of ties and how they relate to each other. The multilevel four-cycle in Figure~\ref{fig:socrecomb}, by itself, represents a form of complementarity. Where $i$ reports that $u$ is connected to $s$, $j$ reports that $u$ is connected to $v$. Considered in combination with the alignment of Figure~\ref{fig:socentrain} we can think of two ways in which to interpret the tie $(j,uv)$. If $i$ and $j$ agree on $\{u,s \}$, we would expect that they would also agree on $\{u,v\}$, and we would expect to see few of the configurations in Figure~\ref{fig:socrecomb}. Similarly, under a social process promoting agreement, we would expect the configuration of Figure~\ref{fig:socrecomb} to be unstable and tend to be \emph{recombined} to the multilevel triangle of Figure~\ref{fig:socentrain}.

\begin{figure}[ht]
\begin{center}
\begin{tikzpicture}
\begin{scope}

\boldmath
\tikzstyle{every
node}=[shape=circle,minimum size=2ex]
\tikzstyle{every path}=[very thick, -stealth', shorten <=2pt,
shorten >=2pt, -]

\node[draw] (ismall) at (0,0)  {{\small $i$}} ;
\node[draw] (jsmall) at (2,2)  {{\small $j$}} ;
\node[draw] (ksmall) at (4,0)  {{\small $k$}} ;
\node[draw,shape=rectangle,rounded corners]  (uvsmall) at (2,4) {{\small $uv$}};
\node  (a) at (0,3.5) {{\small (a)}};
\draw[very thick]  (ismall) -- (jsmall);\draw[very thick]  (jsmall) -- (ksmall);\draw[very thick]  (ismall) -- (ksmall);
\draw[very thick]  (jsmall) -- (uvsmall);

\node[draw] (ismall2) at (0,6)  {{}} ;
\node[draw] (ksmall2) at (4,6)  {{}} ;
\node[draw,shape=rectangle,rounded corners]  (ussmall) at (2,8) {{\small $us$}};
\node  (b) at (0,7.5) {{\small (b)}};
\draw[very thick]  (ismall2) -- (ksmall2);\draw[very thick]  (ismall2) -- (ussmall);\draw[very thick]  (ksmall2) -- (ussmall);

\node[draw] (i) at (8,4)  {{$i$}} ;
\node[draw] (j) at (10,2)  {{$j$}} ;
\node[draw] (k) at (12,4)  {{$k$}} ;
\node[draw,shape=rectangle,rounded corners]  (uv) at (10,0) {{\small $uv$}};
\node[draw,shape=rectangle,rounded corners]  (us) at (10,6) {{\small $us$}};
\node  (c) at (8,5.5) {{\small (c)}};
\draw[red, very thick, dashed]  (j) -- (us);  \draw[red, very thick, dashed]  (j) -- (uv);  
 
\draw[very thick]  (i) -- (j);
\draw[very thick]  (i) -- (k);
\draw[very thick]  (k) -- (j);
\draw[very thick]  (i) -- (us);
\draw[very thick]  (k) -- (us);

\node  (minus) at (10.5,1) {{\small $(-)$}};
\node  (plus) at (10.5,4.5) {{\small $(+)$}};


\end{scope}
\end{tikzpicture}

\end{center}
\caption{ Capturing triadic pressure to conform. If there in addition to the pairwise conformity induced by (b) exists triadic pressure to confirm, we expect (a) to be rare as the combination of (a) and (b) induces extra-dyadic tendency for $j$ to prefer $us$ to $uv$ in (c)}
\label{conf:EXTA}
\end{figure}
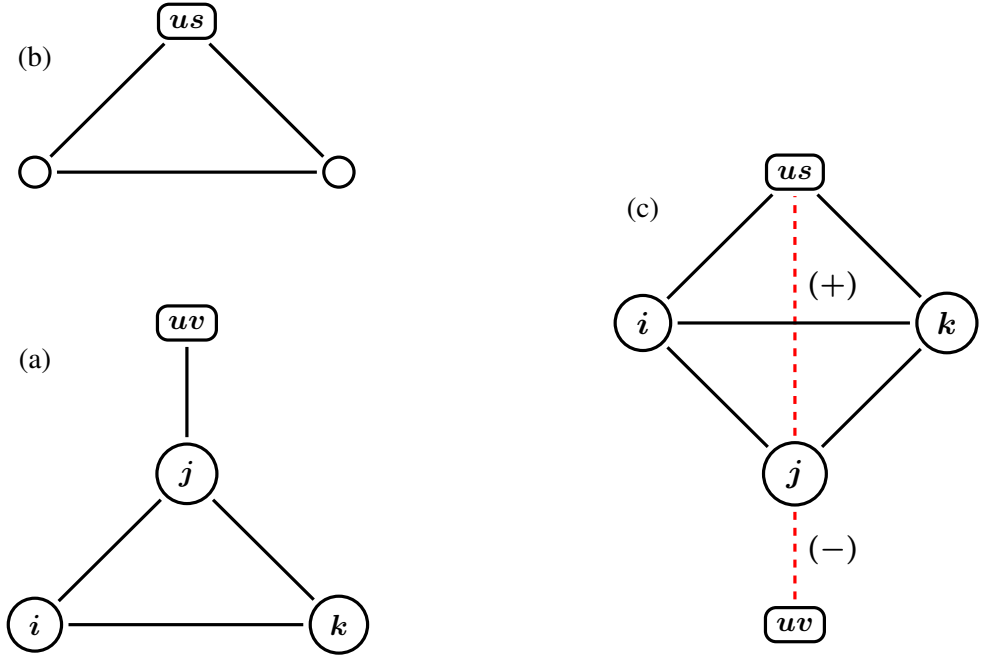

An example of extra-dyadic dependencies in $\mathcal{M}$ and their effect on $\mathcal{H}$ could be that if three nodes $i$, $j$, and $k$ are a clique in $\mathcal{M}$, then they are more likely to agree on $\mathcal{H}$ than what we would see as a result of the dyadic agreement of Figure~\ref{fig:socentrain}. In Figure~\ref{conf:EXTA}, three dyads agreeing on $\{ u,s \}$ would be the additive effect of a positive tendency for configuration (b). If there is triadic pressure over and above this additive dyadic pressure, then you would expect that there would be a tendency against $j$ to in addition nominate $\{ u,v \}$ (Figure~\ref{conf:EXTA}(a)). Put together, we would expect the tie $(j,us)$ to be more likely than $(j,uv)$ in Figure~\ref{conf:EXTA}(c).

\begin{figure}[ht]
\begin{center}
\begin{tikzpicture}
\begin{scope}

\boldmath
\tikzstyle{every
node}=[shape=circle,minimum size=2ex]
\tikzstyle{every path}=[very thick, -stealth', shorten <=2pt,
shorten >=2pt, -]
\node (Hi) at (0,5)  {{$\mathcal{H}_i$}} ;
\node[draw] (ui) at (0,2)  {{$u$}} ;
\node[draw] (si) at (0,4)  {{$s$}} ;
\node (ia) at (0,1)  {{$i$}} ;

\node (yij) at (2,0)  {{$y_{ij}=1$}} ;

\node (Hj) at (4,5)  {{$\mathcal{H}_j$}} ;
\node[draw] (uj) at (4,2)  {{$u$}} ;
\node[draw] (sj) at (4,4)  {{$s$}} ;

\node (ja) at (4,1)  {{$j$}} ;


\draw[very thick]  (ui) -- (si); 
\draw[very thick]  (uj) -- (sj);

\node[draw] (i) at (8,2)  {{$i$}} ;
\node[draw] (j) at (10,2)  {{$j$}} ;

\node[shape=rectangle,rounded corners,draw]  (us) at (9,4) {{\small $us$}};
  \node (C) at (9,5)  {{$\mathcal{C}$}} ;
  
\draw[very thick, -stealth', shorten <=2pt, shorten >=2pt, ->]  (i) -> (us);  
\draw[very thick, -stealth', shorten <=2pt, shorten >=2pt, ->]  (j) -> (us); 
\draw[very thick]  (i) -- (j); 


\end{scope}
\end{tikzpicture}

\end{center}
\caption{ Alignment of ties of $\mathcal{H}_i$ and $\mathcal{H}_j$ for $i$ and $j$ with $\{i,j\} \in \mathcal{M}$,  expressed as a multilevel network of networks. A balanced configuration that representing homophily, social influence, or social selection. For example, the elements of my belief system are the same as those of my friends.}
\label{fig:socentrain}
\end{figure}
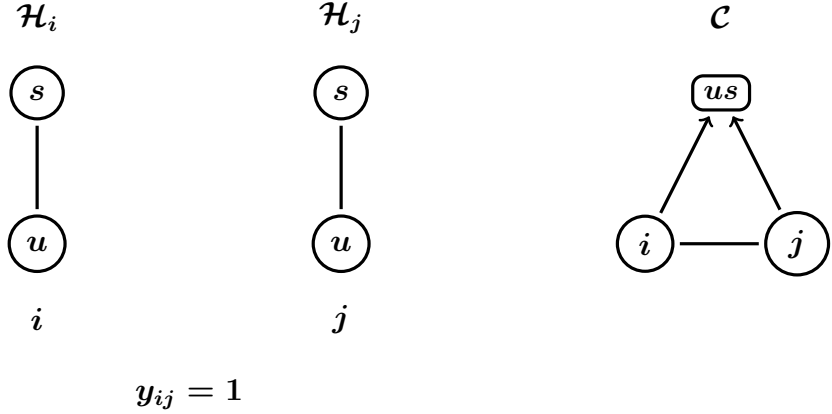

\begin{figure}[ht]
\begin{center}
\begin{tikzpicture}
\begin{scope}

\boldmath
\tikzstyle{every
node}=[shape=circle,minimum size=2ex]
\tikzstyle{every path}=[very thick, -stealth', shorten <=2pt,
shorten >=2pt, -]
\node (Hi) at (1,5)  {{$\mathcal{H}_i$}} ;
\node[draw] (ui) at (1,2)  {{$u$}} ;
\node[draw] (si) at (0,4)  {{$s$}} ;
\node[draw] (vi) at (2,4)  {{$v$}} ;
\node (ia) at (1,1)  {{$i$}} ;

\node (yij) at (3.5,0)  {{$y_{ij}=1$}} ;

\node (Hj) at (5,5)  {{$\mathcal{H}_j$}} ;
\node[draw] (uj) at (5,2)  {{$u$}} ;
\node[draw] (sj) at (4,4)  {{$s$}} ;
\node[draw] (vj) at (6,4)  {{$v$}} ;

\node (ja) at (5,1)  {{$j$}} ;

\draw[very thick]  (ui) -- (si); 
\draw[very thick]  (uj) -- (vj);

\node[draw] (i) at (8,2)  {{$i$}} ;
\node[draw] (j) at (10,2)  {{$j$}} ;

\node[shape=rectangle,rounded corners,draw]  (us) at (8,4) {{\small $us$}};
\node[shape=rectangle,rounded corners,draw]  (uv) at (10,4) {{\small $uv$}};
  \node (C) at (9,5)  {{$\mathcal{C}$}} ;
  
\draw[very thick, -stealth', shorten <=2pt, shorten >=2pt, ->]  (i) -> (us);  
\draw[very thick, -stealth', shorten <=2pt, shorten >=2pt, ->]  (j) -> (uv); 
\draw[red, very thick]  (uv) -- (us);  
\draw[very thick]  (i) -- (j); 


\end{scope}
\end{tikzpicture}

\end{center}
\caption{ Complementarity of ties of $\mathcal{H}_i$ and $\mathcal{H}_j$ for $i$ and $j$ with $\{i,j\} \in \mathcal{M}$,  expressed as a multilevel network of networks. This is an unbalanced configuration if friends $i$ and $j$ disagrees on the role of $u$ or a transient configuration using the same sign $u$ differently could lead to alignment.}
\label{fig:socrecomb}
\end{figure}
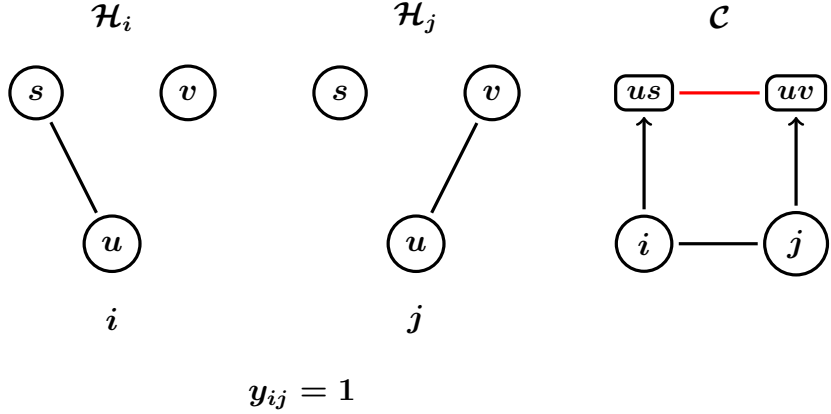

\subsubsection{Alignment with criterion graph}
The representation of $\mathcal{H}_1,\ldots,\mathcal{H}_n$ in terms of a multilevel network $\mathcal{C}$ is already a representation of a \emph{massively multiplex} network. This multiplex network may also be modelled jointly with the social network $\mathcal{M}$. In addition we may consider the alignment of $\mathcal{H}_1,\ldots,\mathcal{H}_n$ with a criterion graph $\mathcal{G}$. In the first instance we may consider basic entrainment of $\mathcal{H}$ and $\mathcal{G}$. The simplest form of alignment may be translated as in Figure~\ref{CovERGMAlignment}, where the grey node indicates that $A_{uv}=1$ for $\pi(u,v)=r$ means that $D_r=1$.

\begin{figure}[ht]
\begin{center}
\begin{tikzpicture}
\begin{scope}

\boldmath
\tikzstyle{every
node}=[shape=circle,draw,minimum size=2ex]
\tikzstyle{every path}=[very thick, -stealth', shorten <=2pt,
shorten >=2pt, -]
\node (u) at (0,2)  {{$u$}} ;
\node (s) at (2,4)  {{$s$}} ;
\node (v) at (4,2)  {{$v$}} ;
\node (i) at (8,2)  {{$i$}} ;

\node[shape=rectangle,rounded corners]  (us) at (6,4) {{\small $us$}};
\node[shape=rectangle,rounded corners,fill=gray]  (uv) at (8,6) {{\small $uv$}};
\node[shape=rectangle,rounded corners]  (sv) at (10,4) {{\small $sv$}};
\draw[very thick]  (u) -- (s);  \draw[very thick]  (s) -- (v);  
\draw[red, very thick, dashed]  (u) -- (v);
\draw[very thick, -stealth', shorten <=2pt, shorten >=2pt, ->]  (i) -> (us);  
\draw[very thick, -stealth', shorten <=2pt, shorten >=2pt, ->]  (i) -> (sv);

\draw[red, very thick]  (uv) -- (us);  \draw[red, very thick]  (sv) -- (us);  
\draw[red, very thick]  (uv) to (sv);


\end{scope}
\end{tikzpicture}

\end{center}
\caption{ Mapping closure in $\mathcal{G}$  of a two-path in  $\mathcal{H}_i$  to multilevel configuration in $\mathcal{C}$ where $A_{uv}=1$ is represented by the attribute $D_{r}=1$ (grey), for $\pi(u,v)=r$. Do informants indirectly link, for example, concepts that are directly linked in a reference graph?}
\label{CovERGMClosure2}
\end{figure}
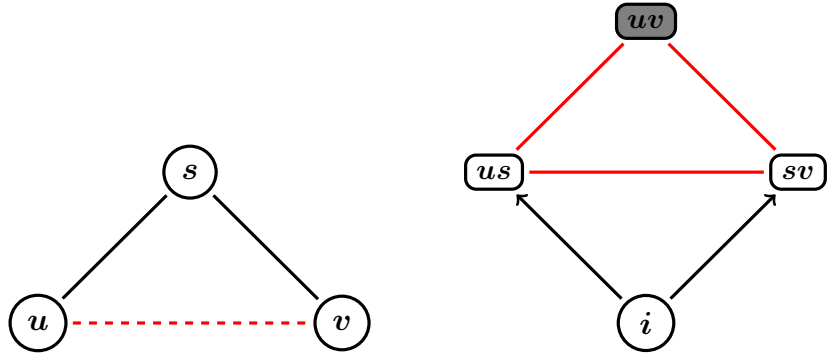

Other multiplex configurations may be expressed in terms of various combinations of ties in $\mathcal{C}$ and attributes $\mathbf{D}$. If there is a tendency for $i$ to directly connect $u$ and $v$ ($u,v \in \mathcal{N}$) that are indirectly connected in $\mathcal{G}$, this is represented as a the triangle-edge configuration in Figure~\ref{CovERGMClosure}. If there is a tendency for $i$ to indirectly connect $u$ and $v$ ($u,v \in \mathcal{N}$) that are directly connected in $\mathcal{G}$, this is represented as a the multilevel four-cycle with a cord in Figure~\ref{CovERGMClosure2}.

\begin{figure}[ht]
\begin{center}
\begin{tikzpicture}
\begin{scope}

\boldmath
\tikzstyle{every
node}=[shape=circle,draw,minimum size=2ex]
\tikzstyle{every path}=[very thick, -stealth', shorten <=2pt,
shorten >=2pt, -]
\node (u) at (0,2)  {{$u$}} ;
\node (v) at (2,2)  {{$v$}} ;
\draw[red, very thick,dashed]  (u) to [bend left] (v);  
\draw[very thick]  (u) to [bend right] (v);

\node[shape=rectangle,rounded corners,fill=gray]  (uv) at (4,2) {{\small $uv$}};
\node (i) at (4,0)  {{$i$}} ;
\draw[very thick]  (i) -- (uv);  

\end{scope}
\end{tikzpicture}

\end{center}
\caption{ Mapping alignment of $\mathcal{H}_i$ (black solid) and $\mathcal{G}$ (red dashed) to the configuration in $\mathcal{C}$ where $A_{uv}=1$ is represented by the attribute $D_{r}=1$, for $\pi(u,v)=r$. If the criterion graph represents the ground truth of the relationship on $\mathcal{N}$, the extent of alignment corresponds to the accuracy of informants. }
\label{CovERGMAlignment}
\end{figure}
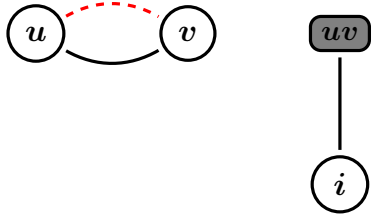

\begin{figure}[ht]
\begin{center}
\begin{tikzpicture}
\begin{scope}

\boldmath
\tikzstyle{every
node}=[shape=circle,draw,minimum size=2ex]
\tikzstyle{every path}=[very thick, -stealth', shorten <=2pt,
shorten >=2pt, -]
\node (u) at (0,2)  {{$u$}} ;
\node (s) at (2,4)  {{$s$}} ;
\node (v) at (4,2)  {{$v$}} ;
\node (i) at (8,2)  {{$i$}} ;

\node[shape=rectangle,rounded corners,fill=gray]  (us) at (6,6) {{\small $us$}};
\node[shape=rectangle,rounded corners]  (uv) at (8,4) {{\small $uv$}};
\node[shape=rectangle,rounded corners,fill=gray]  (sv) at (10,6) {{\small $sv$}};
\draw[red, very thick, dashed]  (u) -- (s);  \draw[red, very thick, dashed]  (s) -- (v);  
\draw[very thick]  (u) -- (v);
\draw[very thick, -stealth', shorten <=2pt, shorten >=2pt, ->]  (i) -> (uv);

\draw[red, very thick]  (uv) -- (us);  \draw[red, very thick]  (sv) -- (us);  
\draw[red, very thick]  (uv) to (sv);


\end{scope}
\end{tikzpicture}

\end{center}
\caption{ Mapping closure in $\mathcal{H}_i$ (black solid)   of a two-path in $\mathcal{G}$ (red dashed) to multilevel configuration in $\mathbf{C}$ where $A_{uv}=1$ is represented by the attribute $D_{r}=1$ (grey), for $\pi(u,v)=r$. When the criterion graph represents a true or normative relationship among $\mathcal{N}$, informant $i$ might associate nodes $u$ and $v$ whenever they are actually only indirectly connected.}
\label{CovERGMClosure}
\end{figure}
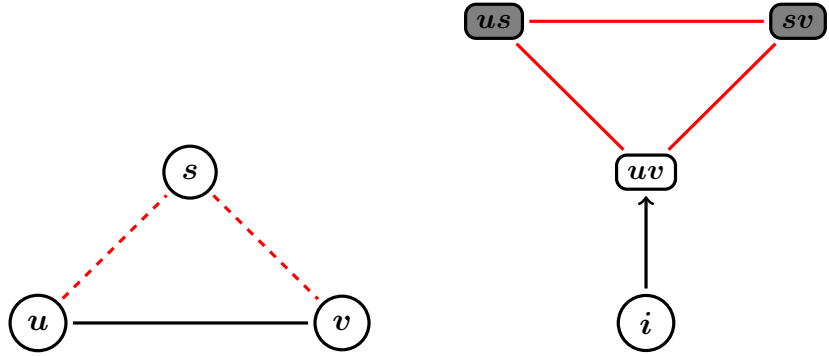

\subsubsection{Connecting layers}
In sociomaterial networks (Contractor et al., 2011), sociosemantic networks (Basov, 2020), and socioecological networks (Bodin et al., 2016), a social network amongst the nodes of $V$ is connected to a network on $\mathcal{N}$ through some form of two-mode ties in a multilevel network (Lazega et al., 2008). Some of these two-mode ties may lend themselves more readily to be represented as networks $\mathcal{H}_i$, such as for example consumer ($V$) preferences amongst products ($\mathcal{N}$), rather than a consumer by product, two-mode network (Wang et al., 2015). In the multilevel semantic network of Basov (2020) people $V$ are connected to concepts $\mathcal{N}$ through usage of concepts, represented by a two-mode network. The network amongst $\mathcal{N}$ is however aggregated from the individual semantic networks that are essentially $\mathcal{H}_1,\ldots,\mathcal{H}_n$.



\subsubsection{Computational considerations\label{sec:compcons}}
Having translated the original data $ \mathcal{H}$, $  \mathcal{G}$, and $ \mathcal{M}$  to a multilevel network, the dependencies between the different types of ties lend themselves to investigation using for example multilevel ERGM (Peng et al., 2013), assuming that the criterion graph $\mathcal{G}$ is fixed and treated as an explanatory network. If, as in CSS, $\mathcal{G}$ is unobserved, the representation applies only either conditionally on an assumed criterion graph or in the absence of an assumed criterion graph. In principle, $\mathcal{G}$ could be treated as a latent, unobserved attributes (e.g. Koskinen, 2009; Schweinberger, 2019) but this would most likely be practically infeasible.

For more general methods for analysing multilevel networks, Hollway, Lomi, Pallotti, and Stadtfeld (2017) redefine the multilevel networks as a blocked one-mode network with structural zeros preventing ties in layers of the network where some ties are not defined (this is further described in Snijders, 2019). If the aim of the analysis is not to model all types of ties but rather focus on one set of ties, using the others as predictors, multilevel configurations can be projected to dyadic covariates, for example for modelling a social network conditional on affiliations $\mathcal{W}$ and top-level networks $\mathcal{Q}$. For example, assume that we want to model $\mathbf{Y}$ conditionally on $\mathbf{W}$ and $\mathbf{Q}$, then the statistic for the multilevel four-cycle in Figure~\ref{fig:socrecomb}, $Y_{ij}\sum_{r,t}W_{ir}W_{jt}Q_{rt}$, can be represented as a dyadic covariate $h_{ij}=\sum_{r,t}W_{ir}W_{jt}Q_{rt}$ for $Y_{ij}$ (Stys et al., 2020).  

Assuming that we are using the entirety of $\mathcal{E}$, the number $e$ of `top-level' nodes in the multilevel representation will be too large for the translations to be practically feasible for large $N$. Consider for example the size $e$ for a CSS for a standard size network. Careful consideration therefore has to be given to what defines meaningful subsets of $\mathcal{E}$. The subset of $\mathcal{E}$ used for the top-level does not have to be the induced set of possible relations on a subset of $\mathcal{N}$. We may allow for some nodes in $\mathcal{N}$ to be represented in more pairs than others. In fact, we may let $\mathcal{Q}$ be the line-graph of some graph on $\mathcal{N}$ that is, say, the union of the graphs $\mathcal{H}_1,\ldots,\mathcal{H}_n$. We will show an example of this next, in our empirical illustration.

\section{Illustration of social production of knowledge}
\subsection{Data}
Our data is obtained from a pilot study of a larger project (`Creation of knowledge on ecological hazards in Russian and European local communities') that aims at investigating how knowledge in local flood-prone communities conforms to knowledge of experts (i.e. flood management agencies and authorities). The set $\mathcal{N}$ consists of concepts or \emph{signs}, relevant for flood management.

The signs chosen by experts is a graph $\mathcal{G}$ on nodes $\mathcal{N} \subset \Omega$ with edge set $E \subset \binom{\mathcal{N} }{2}$. The signs $\mathcal{N} $ thus constitute the expert vocabulary. We denote the adjacency matrix of $\mathcal{G} $ by $\mathbf{A}$. Throughout we assume that $\mathbf{A}$ is fixed, exogenous, and unchanging. In terms of elements
\begin{equation*}  
	A_{uv} = \left\{
	\begin{array}{lr}
		1,&\text{if sign } u \text{ is connected to sign } v \\
		0,&\text{otherwise }\\
	\end{array} 
	\right. 
\end{equation*}
The expert semantic network has been constructed using `UDPipe' (Wijffels et al., 2019) applied to expert texts - flood management-related documents issued by flood risk management agencies and authorities.

In the pilot, fieldwork in a site for local flood management has yielded the social network and the individual semantic networks for $n=15$ individuals - members of two local flood groups voluntarily involved in flood management. For each $i \in V$, the individual semantic structures (Basov and de Nooy, 2019) were constructed using the same software applied to transcripts of semi-structured interviews with the members, as 
\begin{equation*}  
	X_{iuv} = \left\{
	\begin{array}{lr}
		1,&\text{if } i \text{ nominates a tie between } u  \text{ and } v \\
		0,&\text{if } i \text{ does not nominate a tie between  } u  \text{ and } v \\
	\end{array} 
	\right. 
\end{equation*}
Here we focus the analysis on a subset of $\mathcal{E}$ with $e=634$ motivated by previous work on multilevel socio-semantic networks where the individual semantic structures are aggregated into a local knowledge network wherein ties exist between concepts that are used together by at least two individuals (Basov, 2020). For the model, this means that we only model tie presence between pairs of concepts that are present in the (implicit) local knowledge network, and thus exclude from the analysis those pairs which are never used together in the local context. The node set is thus the nodes of the line graph of the graph that is the union of expert and the individual semantic networks.  This vastly reduces the computational cost as the size of the top-level of the network is reduced from being in the region of tens of thousands to a manageable, and interpretable size. Furthermore, $\mathcal{W}$ does not have any isolate  $\mathcal{P}$ nodes.

The social network is an undirected social network among the actors in $V$ with the adjacency matrix $\mathbf{Y}$ defined as having elements
\begin{equation*}  
	Y_{ij} = \left\{
	\begin{array}{lr}
		1,&\text{if } i \text{ nominates } j  \\
		0,&\text{ otherwise }\\
	\end{array} 
	\right. 
\end{equation*}
The social network has been derived using visually verified sociometric surveys, triangulated with interview and observational data  to guarantee high quality data (for details on the procedure, see Basov, 2020).

We will analyse the dataset using MPNet and will refer to the corresponding effect names in the description of the configurations we investigate (Wang et al., 2014). The logic of the effect names, generally speaking, is that the type of structure, e.g. `triangle', is followed by letters referring to what networks are involved, where X is the bipartite, cross-level network, A is the `bottom' network, and B is the `top' network. In our example, the social network (SN) is the `bottom' network A; the individual semantic networks or local meaning structures the bipartite network X; and $\mathcal{Q}$ is the `top' network B.


\subsection{Social construction of knowledge}
In the first instance, in the flood management groups studied, we aim to investigate how the actors $V$ negotiate meaning, irrespective of how flood management signs are related in the `normative' expert network. Symbolic interactionist conception of social behaviour is based on the premise that people collectively construct knowledge about reality rather than passively reproduce images of the world imposed on them:
\begin{quote}
$\ldots$ [h]uman group life on the level of symbolic interaction is a vast process in which people are forming, sustaining, and transforming the objects of their world as they come to give meaning to objects. (Blumer, 1986: 12; Mead, 1934)
\end{quote}
Interacting ($\mathcal{M}$), individuals use signs and associations of signs ($\mathcal{H}$) that refer to a context where their interaction unfolds and constitute knowledge about reality (Mead, 1934). The interaction context here is flood management in their communities. The act of knowledge creation can be thought of as three interrelated processes that lead to multilevel, social entrainment as in Figure~\ref{fig:socentrain} (TriangleXAC). Firstly, to indicate the meaning of an object/action actor A would use signs familiar to B and avoid using signs that are incomprehensible for the group that both A and B belong to (e.g. technical jargon).
This entails two levels of entrainment, one is the entrainment in Figure~\ref{fig:basicentrain} (XASB) and the more structural entrainment in Figure~\ref{fig:relatedness} (XACA), neither of which involves defining the interaction in terms of $\mathcal{M}$, rather using membership in the group $V$ as the reference point. 
Secondly, in interaction between A and B meanings indicated by A are either confirmed or rejected by B. This constitutes the other level of entrainment presented in 
Figure~\ref{fig:socentrain}. The shared signs and meaningful associations between them constitute community knowledge. Thirdly, the confirmation of the indicated meanings by actor B becomes a stimulus for individual A to continue interaction.
According to these three processes, actors can reproduce and transform local knowledge. Reproduction of knowledge involves actors restating already shared signs and associations between signs . Transformation involves reconfiguration of associations between signs as a result of interaction between actors. Knowledge reproduction and transformation are carried out through several mechanisms working at dyadic and extra-dyadic levels and summarised by Antonyuk et al (2019). The cross-sectional data, however, do not allow distinguishing between knowledge reproduction and transformation.   Similarly, for the social ties, they may be created from various forms of social alignment or they may be reconfigured to reduce breach of social alignment. In what follows we offer examples of such mechanisms and their statistical representations.



\subsubsection{Knowledge reproduction and transformation at dyadic level: Selection and contagion}
Knowledge reproduction can take place through the mechanism of `selection'. This mechanism can be observed in different contexts, for example, when people seek to establish reliable knowledge about an object or an event. In this situation, the identical elements in different indicated meanings are considered reliable, while the divergent parts are dismissed as subjective or inaccurate.

For example, discussing flood relief measures, A and  B may disagree on what kind of help, e.g. money or material supplies,  needs to be  provided to the victims of a recent flood . At the same time, disagreeing on a particular type of aid throughout interaction, they at least agree  that some help has to be provided to the victims (both A and B retain links between signs `provide' and `victim'). Thus, by dismissing conflicting meanings (i.e. dismissing links `victim' - `money' and `victim' - `material supplies' previously used by A and B, respectively), the mechanism of selection helps reproduce the common ground necessary for collaboration between the actors. 

Knowledge transformation can take place through the mechanism of `contagion'. Contagion occurs when one group member starts using an association between signs as a result of interacting with another group member using them (Burt 1987; Monge and Contractor 2003; Carley 1986; Coleman 1988). As a result, the association becomes shared and thus becomes part of local group's knowledge. For example, if A introduces the idea of monitoring water levels at the local river to B, who is unfamiliar with this idea, B may adopt the association between existing signs `water-level' - `monitoring' from A, that as a result becomes part of group knowledge. 
Knowledge transformation may also happen when there emerges a new problem  not captured by previously existing group knowledge (e.g. flood water cannot be eliminated with a pump) requiring recombination of existing signs (Bolton 1981; Hollander \& Gordon 2006; Etzrodt 2008). Reflecting on existing approaches to the problem, actor A  can come up with an idea of a better flood protection device combining existing devices in a novel way (e.g., a physical barrier combined with a pump that automatically removes water). The new idea would reveal itself in a recombination of corresponding signs and emergence of a new association `barrier' - `pump'. Actor B may support the new idea and adopt the association between signs `barrier'- `pump'. Structurally this will be represented as an  unstable four-cycle in Figure~\ref{fig:socrecomb}. B may adopt the new association and dismiss the old one - between `water' and `pump' (the four-cycle) - in favour of the socially aligned Figure~\ref{fig:socentrain}. This \emph{recombination} would suggest a positive effect for the multilevel triangle of the type Figure~\ref{fig:socentrain} (TriangleXAC) and a negative effect for the four-cycle of the type Figure~\ref{fig:socrecomb} (C4AXB).

\subsubsection{Knowledge reproduction and transformation at extra-dyadic level}
At the extra-dyadic level, transformation of knowledge may occur through the mechanism of preferential attachment, when actors `frame' an element of shared knowledge by adopting associations with signs that already have general, often emotionally charged meanings (Schultz et al., 2012; Snow et al., 1986), rather than with signs that denote more specific meanings. Signs used for `framing' usually have many shared associations with other signs because of their generality or abstractness that allows them to enter many different contexts. For instance, actor A may argue that the group should adopt a prevention approach to `fluvial' floods (that is, floods caused by excessive water in a river). The new association between existing signs used by A, e.g., `flood prevention', invokes the `prevention' frame that is known for actors in group $V$ from other contexts like health or road safety. Therefore, the association `flood'- `prevention' is likely to be adopted by another group member, B (and hence, become part of group knowledge), unlike a more specialised association `fluvial floods' that does not involve any frame previously known to member B. A combination of configurations that is consistent with this preferential attachment is not to select isolated ties $\{ s,u \}$ (negative XASB), nor are ties $\{ s,u \}$ selected merely because there are many pairs with $s$ (negative EXTB). A concept pair $\{s,u \}$ if other ties of $s$ are reported, captured through the configuration in Figure~\ref{fig:spillover} (TriangleXBX) (positive), and if many socially tied individuals chose the same $\{ s,u \}$ (positive effect for the configuration in Figure~\ref{fig:socentrain}, TriangleXAX) not ties $\{s,v\}$ that also include $s$ (tendency against configurations Figure~\ref{fig:socrecomb}, negative C4AXB).

In relational patterns involving three individuals, knowledge may be reproduced through the mechanism of triadic pressure. A triad embodies supra-individuality (Simmel, 1950: 257) and downplays individuality, diminishing the power of a single actor to determine the outcome of the whole interaction process. Relations in triads are `less free, less independent, more constrained' than in dyads (Krackhardt, 1999: 185). In addition, actors need 
\begin{quote}
[$\ldots$] social affirmation or reinforcement from multiple sources [$\ldots$] since contact with a single active neighbour is not enough to trigger adoption (Centola and Macy, 2007: 705).
\end{quote}
Of course, the difference between total and average exposure on the probability of adoption is a key consideration in diffusion models (see Strang and Tuma, 1993, and related work) as well as in influence models (Leenders, 2002). Therefore, in triads, reproduction of existing shared knowledge is more efficient than in a dyad (Krackhardt, 1999): if in a triad two interacting individuals A and B share associations between signs e.g., `flood' - `management', the third individual C interacting with both is likely to start sharing these associations as well. This effect is over and above dyadic contagion, so that the propensity of the third individual to share a sign or an association is higher than if he or she was subjected to contagion by two alters who are not socially tied with each other. This triadic pressure translates into the process described in Figure~\ref{conf:EXTA}(c) as a positive effect for the configuration in Figure~\ref{fig:socentrain} (TriangleXAX) and a negative effect for the configuration in Figure~\ref{conf:EXTA}(a)(EXTA).

\subsubsection{Entrainment with normative knowledge}
The associations between signs in the local knowledge are affected not only  by social relationships between actors but also by expert knowledge imposed on the flood management groups by authorities, that is the normative relations between signs indicated by $\mathcal{G}$. The influence of expert knowledge on local knowledge occurs through several mechanisms that reflect structural changes in the local knowledge network conditional on the structure of $\mathcal{G}$.  Basov and Brennecke (2017) perform a multiplex analysis where an aggregate local knowledge network is compared to the criterion graph $\mathcal{G}$

There are a number of ways in which we can extend the multiplex dependencies of an aggregated local knowledge network on $\mathcal{G}$, to dependencies of $\mathcal{H}$ on $\mathcal{G}$, and examine theoretically-derived mechanisms of expert knowledge influence on local knowledge, e.g. such as those that result in alignment in Figure~\ref{CovERGMAlignment} and the two forms of closure in  Figure~\ref{CovERGMClosure} and Figure~\ref{CovERGMClosure2}. We include a basic entrainment corresponding to a mechanism we call `basic reproduction'. It occurs when a local group starts associating preexisting signs in the same way they are associated by the experts. For example, experts can talk about `rivers' as sources of flood risks while locals may not associate `river' with `risk' at all. Locals may realise that the experts' way of thinking about the river as a source of risks could be useful for them, e.g., to discuss the problem of flooding with authorities. 

Hence, following the experts, they start associating `river' and `risk'. The mechanism is modelled using the effect Expert XEdgeB corresponding to the configuration in Figure~\ref{CovERGMAlignment}. 

To control for the possibility that some concept pairs in $\mathcal{G}$ may be salient only because they involve signs $s$ that are part of many other concept pairs, we include the interaction $\sum_{i,r}W_{ir}D_r\sum_{t \neq r} Q_{rt}$ which is the statistic Expert Star2BX. This statistic corresponds to a mechanism we call `popularity pressure' that describes a situation when a local group starts associating signs that are part of many other associations in the experts' knowledge. For example, speaking about flood risk management, experts may pay significant attention to communities' resilience to flood hazards and highlight the importance of creating flood plans to ensure all stakeholders are prepared for potential floods. Because a local group observes the signs `resilience' and `plan' as focal for experts, they also start associating these two signs when speaking about a local document they produce to be prepared for floods - a local group `resilience plan'. 

 
\subsection{Results}

The results for two models are presented in Tables~\ref{tab:MERGMres}. For Model 1, the social network abides by standard social processes judging by the one-mode effects (Snijders et al., 2006; Lusher et al., 2013), with no heterogeneity in popularity (insignificant ASA) but with evidence for triadic closure (positive ATA or GWESP). The bipartite network is modelled using the three terms XEdge, XASA, and XACA. We will interpret these in relation to the multilevel statistics making up the rest of the table.

In terms of reproduction and transformation of knowledge, the effects go largely in the direction of the predictions. The positive TriangleXAC is consistent with basic contagion or selection. The positive effect, in Model 1, for TriangleXAC and a negative effect for C4AXB is consistent with recombination of concept pairs.

The combination of negative EXTB, positive TriangleXAX, and negative C4AXB, is consistent with the framing operating through preferential attachment.

Amongst the other effects, XACA and TriangleXBX capture a general coherence of the local knowledge structure, connecting concept pairs that are not socially mediated. XEdge, Star2BX, StarAB1X take into account that not all signs are equally represented in $\mathcal{E}$. The statistics Star2BX, L3XBX, and L3AXB also act as lower-order interactions to C4AXB.

Introducing $\mathcal{G}$, we see that there is a significant alignment of the individual semantic networks and the expert network (Expert XEdge). We also see that the dependence on $\mathcal{G}$ completely moderates the effect of C4AXB. This means that some of the recombination and preferential attachment is explained by how concept pairs are related in the expert network. Network-related contagion (TriangleXAX) and local knowledge structure (XACA) are however not affected by the reliance on the expert network.

The effect EXTA, necessary for inferring triadic pressure (see Figure~\ref{conf:EXTA}(a)), is not estimable from this dataset and while the model predicts more configurations EXTA than observed (see GOF in Tables~\ref{tab:MERGMGOF} and \ref{tab:MERGMGOFexpert}), this difference is not significant. In general, the goodness-of-fit is overall acceptable considering the complexity of the data. Some expert-related configurations in the goodness-of-fit suggest that there is scope for investigating more elaborate multiplex effects. Furthermore, in the goodness-of-fit there are higher-order interactions with $\mathcal{Q}$ that could be investigated pursuant theorising in terms of $\mathcal{H}$, $\mathcal{M}$, and $\mathcal{G}$.  Some affiliation configurations are poorly fit by the simple model of Table~\ref{tab:MERGMres}, something which is to be expected for two-mode networks (Wang et al., 2009) and which will be alleviated by incorporating the expert network as a top-level covariate.


		\begin{center}
			\begin{longtable}{p{45mm}rrrrcc}
		\caption{MERGM results for local meaning structures (LMS), social network (SN), and expert semantic network (ESN) (social and bipartite effects). In the Multilevel network diagrams, round nodes are people and square nodes concept-concept pairs. In the network of networks (NoN) diagrams, the semantic networks are labeled with indices in $V$ and where applicable, social ties are indicated by ties between the large circles.}\label{tab:MERGMres}\\
		\hline
		& \multicolumn{2}{|c|}{Model 1}&\multicolumn{2}{|c|}{Model 2}&\multicolumn{2}{|c|}{Representation}\\
		Effects	&	Parameter	&	Stderr	& Parameter	&	Stderr	&	Multilevel	&	NoNs	\\
	\cline{2-7}
	
\endfirsthead
	
\multicolumn{7}{c}%
{{\bfseries \tablename\ \thetable{} -- continued from previous page}} \\
	
\hline
		& \multicolumn{2}{|c|}{Model 1}&\multicolumn{2}{|c|}{Model 2}&\multicolumn{2}{|c|}{Representation}\\
		Effects	&	Parameter	&	SE	& Parameter	&	SE	&	Multilevel	&	NoNs	\\
	\cline{2-7}
	
	\endhead

	\hline \multicolumn{7}{|r|}{{Continued on next page}} \\ \hline
	\endfoot
	
	\hline \hline
	\endlastfoot
	
	\cline{2-7}
	{\small{Density SN (EdgeA)}}	&$	-0.748	$&$	2.019$&$	2.052	$&$	3.245	$& \includegraphics[scale=0.4]{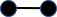}   & \\
	{\small{Centralisation in SN (ASA)}}{\scriptsize{\newline Heterogeneity in the degree distribution of social network}}	&$	-0.861	$&$	0.516	$& $ 	-1.966	$&$	1.125	$& \includegraphics[scale=0.4]{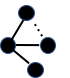}  &\\
	{\small{Triadic closure in SN (ATA)}}{\scriptsize{\newline Geometrically edge-wise shared partners}}	&$	\mathbf{1.331}	$&$	0.48	$& $	\mathbf{1.314}	$&$	0.482	$&\includegraphics[scale=0.4]{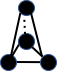} & \\
	{\small{Density LMS (XEdge)}} {\scriptsize{\newline $\sharp$ concept pairs linked}}	&$	-0.772	$&$	0.456	$ &  $	-0.457	$&$	0.473	$  &  \includegraphics[scale=0.4]{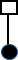} & \includegraphics[scale=0.4]{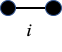}\\
	{\small{Centralisation LMS (XASB)}}{\scriptsize{\newline People in the same context connect the same signs}}&$	\mathbf{-1.075}	$&$	0.27	$ & $	\mathbf{-1.298}	$&$	0.283	$ & \includegraphics[scale=0.4]{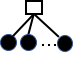} & \includegraphics[scale=0.4]{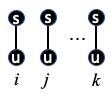}\\
	
{\small{Multiple entrainment LMS (XACA)}} {\scriptsize{\newline Context promotes coordination of sign association}} &$	\mathbf{0.01}	$&$	0.001	$&$		\mathbf{0.009}	$&$	0.001	$  & \includegraphics[scale=0.4]{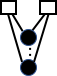} & \includegraphics[scale=0.4]{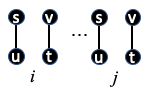}\\
	{\small{Entrainment criterion (Exp. XEdgeB)}}{\scriptsize{\newline Basic reproduction: Signs linked by experts are linked by individuals}} 	& & &  $	\mathbf{0.568}	$&$	0.13	$& \includegraphics[scale=0.4]{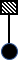} & \includegraphics[scale=0.4]{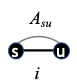} \\

{\small{Assortativity LMS-SN (Star2AX)}}{\scriptsize{\newline Association of centrality SN and density of semantic network}}	&$	-0.039	$&$	0.039	$& $	-0.062	$&$	0.042	$   & \includegraphics[scale=0.4]{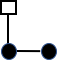} & \includegraphics[scale=0.4]{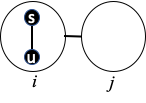}\\
{\small{Balance LMS-SN	(TriangleXAX)}}{\scriptsize{\newline Selection, influence, and recombination}}	&$	\mathbf{0.354}	$&$	0.103	$&  $	\mathbf{0.352}	$&$	0.102	$      & \includegraphics[scale=0.4]{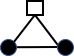} & \includegraphics[scale=0.4]{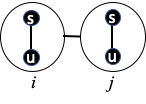}\\
{\small{Key signs (Star2BX)}}{\scriptsize{\newline Are signs used in many meanings connected to many other signs}}	&$	-0.001	$&$	0.126	$&  $	-0.064	$&$	0.135	$      & \includegraphics[scale=0.4]{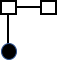} & \includegraphics[scale=0.4]{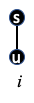}\\
{\small{Key signs (StarAB1X)}}{\scriptsize{\newline Non-linear effect of the number of pairs that includes a sign on its popularity}}	&$	0.058	$&$	0.064	$&   $	0.081	$&$	0.07	$      & \includegraphics[scale=0.4]{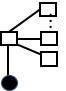} & \includegraphics[scale=0.4]{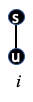}\\
{\small{Centralisation LMS (TriangleXBX)}}{\scriptsize{\newline Centrality of signs across LMS}}	&$	\mathbf{0.065}	$&$	0.006	$&    $	\mathbf{0.065}	$&$	0.006	$       & \includegraphics[scale=0.4]{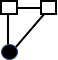} & \includegraphics[scale=0.4]{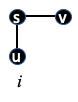}\\
{\small{Complimentary use LMS (L3XBX)}}{\scriptsize{\newline Different usage of the same sign}}	&$	-0.024	$&$	0.015	$&$	-0.027	$&$	0.015	$& \includegraphics[scale=0.4]{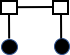} & \includegraphics[scale=0.4]{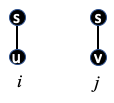}\\
{\small{Preferential attachment LMS (EXTB)}}	&$	\mathbf{-0.001}	$&$	0.000	$&    $	\mathbf{-0.001}	$&$	0.000	$        & \includegraphics[scale=0.4]{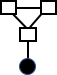} & \includegraphics[scale=0.4]{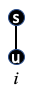}\\
	
{\small{Alignment criterion on salient signs (Exp. Star2BX)}}{\scriptsize{\newline Popularity pressure: Pairs linked by experts that includes frequently used signs}} & &	&$	0.001	$&$	0.001	$&\includegraphics[scale=0.4]{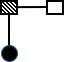} & \includegraphics[scale=0.4]{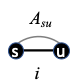}\\
		{\small{Multilayered assortativity (L3AXB)}}{\scriptsize{\newline Association of centrality in SN with concept-pairs involving frequently used signs}}	&$	0.001	$&$	0.001	$&   $	-0.008	$&$	0.005	$       & \includegraphics[scale=0.4]{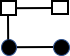} & \includegraphics[scale=0.4]{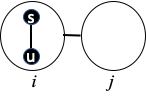}\\
		{\small{Unbalanced four-cycle (C4AXB)}}{\scriptsize{\newline An opportunity to recombine signs to a balanced triad}}	&$	\mathbf{-0.008}	$&$	0.004	$&   $	0.017	$&$	0.014	$        & \includegraphics[scale=0.4]{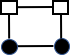} & \includegraphics[scale=0.4]{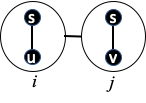}\\
		\hline

		\end{longtable}
		\end{center}

In summary, our analysis of these individual semantic structures amongst the 15 actors in a local flood management groups provides evidence of a variety of mechanisms for socially constructed local knowledge. Some social mechanisms are moderated by the introduction of an external criterion network representing the semantic network of flood management experts. So, while local actors speak about their reality using their own, partially socially constructed, knowledge, this knowledge is not completely independent of the `objective' knowledge of experts. There is further heterogeneity to explore. This may be addressed by developing more elaborate multiplex associations between the individual semantic networks and the expert network. We may also elaborate individual-level effects, for example using actor attributes. Socially dependent heterogeneity could also be explored. Here, the configuration Star2AX captures the association between social popularity and number of concept pairs nominated, but it does not inform us whether people agree with central people. It is not straightforward to see whether a person who is central in the social network tends to be influential. Future research could study the dependence of the overall structure on particular actors, something that would require multilevel elaborations of outlier diagnostics (Koskinen et al., 2018). 

\section{Summary}
We have proposed a conceptual framework for joint analysis of multiple reports on a network, how these relate to a criterion graph, and how a network among reports induces dependencies. The approach rests on transforming the original data into a multilevel network (Lazega et al., 2008) representation. We have discussed the multilevel representation in terms of meaningful, multilevel network configurations that are amenable to empirical investigation using multilevel ERGM (Wang et al., 2013). The representation is however agnostic to the actual analysis method and in the proposed format, networks of networks are amenable to investigation in any analysis framework for multilevel networks, such as stochastic actor-oriented models (Hollway et al., 2017) and blockmodels (\v{Z}ibena and Lazega, 2016).

We provided an illustrative application of the framework to a dataset on semantic networks of socially tied individuals. The translation of data from the original domain into the multilevel representation is not dependent on the specific content of the network data and we have suggested how the approach might be applied in other contexts. Examples include sociosemantic networks (Basov, 2020), socioecological networks (Bodin et al., 2016), and sociomaterial networks (Contractor et al., 2011; Basov, 2018), all of which are explicitly multilevel. Other examples include data collection paradigms that assume repeated observations on networks such as multiplex networks and cognitive social structures. Yet another class would be networks that could be repeatedly observed for units (e.g. intersectional flows in different countries, Leoncini et al., 1996; and, symptom networks, Borsboom and Cramer, 2013) where work has already been done on analysing the networks of these units (social networks in psychiatry, Moreno, 1934; and, economic ties between countries, Squartini et al., 2011; Koskinen and Lomi, 2013). All of these different contexts present unique challenges in specifying meaningful configurations, some of which we have discussed.

While we have detailed some of the properties of the network of networks framework, there are many more that warrant further investigation. There are considerable computational challenges associated with the size of $\mathcal{E}$ and an open question is what different inferences you get when you use all of $\mathcal{E}$, rather than a subset as we did in our empirical example. On the one hand, if $\mathcal{Q}$ is the line graph of a meaningful graph on $\mathcal{N}$, then the line graph preserves all the structure of that graph, but you also impose constraints and reduce the data provided by the informants $i \in V$. Due to the various considerations for our empirical example, code for transforming raw data into the multilevel network representation had to be customised but maybe there is scope for automated routines? We have presented a selection of meaningful local metrics for networks of networks that correspond to specific interdependencies but we have not addressed more general metrics and what they might mean, metrics such as centrality and indices of clustering. In some applications, data would not conform straightforwardly to the canonical case which we have based our exposition on. For example, inter-organisational networks are often elicited from respondents that are members of the organisations of interest, even if there might exist a criterion graph, $\mathcal{G}$, corresponding to formal agreements or financtial relations. If you have more than one respondent per organisation, your raw data would essentially be a network of networks, if you, in addition to the reports on the inter-organisational, say, collaborations, also had an `objective' network amongst informants. However, it is likely that the informants would only report on the collaborations of \emph{their} organisation, in which case each report $\mathcal{H}_i$ would only be a partial network, relating to the ties of $i \in \mathcal{N}$. Furthermore, in most applications you would expect to have a lot of information about the informants and there are many potential ways in which you could incorporate actor attributes, other than as predictors in $\mathcal{M}$. Finally, our list of configurations is by no means exhaustive, and different contexts would likely warrant different types of configurations and explicating what these mean. For example, for two informants $i$ and $j$, if $i$ reports $\{s,u\}$ and $\{s,v\}$, but $j$ directly connects $u$ and $v$, will a social tie between $i$ and $j$ lead $i$ to align with $j$ or $j$ to align their view with $i$, or will they merge their views and decide that $s$,$u$, and $v$ forms a closed triangle?

\makeatletter
\renewcommand\@biblabel[1]{}
\makeatother

\appendix
\section{Glossary}

\begin{table}
{\small{
\caption{\label{tab:notation}Glossary of terms}

\begin{center}
\begin{tabular}{l c l p{65mm}}
\hline

Symbol	& Def	& Description	&	Example	\\

$\mathcal{N}$ & $\{ 1,\ldots,N \}$ & Fundamental set of nodes & Words a semantic network \newline Set of products in a sociomaterial network \newline Forests in a socioecological networks\\
$\mathcal{E}$ & $ {\mathcal{N} \choose 2}$ & Un-ordered pairs of nodes of $\mathcal{N}$ &  Associations between concepts in semantic network \newline Similarities between products in sociomaterial network \newline Seeding between forests in socioecological network \\

$e$ & $e= |\mathcal{E}|$ & Number of possible ties & \\

$V$ & $\{ 1,\ldots,n \}$ & Index of observations on $\mathcal{H}$ & Set of raters (people) \newline Set of observations on a network\\
$\mathcal{H}_i$ & $\mathcal{H}_i(\mathcal{N},E_i)$ & Graph on $\mathcal{N}$ observed for $i$ & The semantic network of $i$ \newline A network of products in country $i$\newline The social network according to $i$\\

$E_i$ & $E_i \subset\mathcal{E}$  & Set of ties in $\mathcal{H}_i$ & A realisation of ties \\

$\mathcal{H}$ & $\{\mathcal{H}_1,\ldots,\mathcal{H}_n \}$ & Collection of observations & Repeated observations on network \newline Multiplex network\\

$ \mathbf{X}_i$ & $(X_{iuv})_{uv \in \mathcal{E}}$ & Adjacency matrix of $\mathcal{H}_i$ & $X_{iuv}=1$ if $\{u,v\} \in E_i$\\

$ \mathcal{G}$ & $\mathcal{G}(\mathcal{N},A)$ & A criterion graph on $\mathcal{N}$ & A externally given semantic network of experts \newline Objective measures on social interaction \\

$ \mathbf{A}$ & & The adjacency matrix of $ \mathcal{G}$ & \\

$ \mathcal{M}$  &  $\mathcal{M}(V,T)$ & A network on the index set $V$ & Social network among raters \newline Trade network among countries \newline Alliances among clans \\

$ \mathbf{Y}$ &  & The adjacency matrix of $\mathcal{M}$ & \\


$\pi$ & $\pi:\mathcal{E}\rightarrow \mathcal{P} $ & Relabelling of tie-variables &\\

$\mathcal{P}$ && Image of $\mathcal{E}$ under $\pi$ & \\

$\mathcal{M}$ && Two-mode graph on $V$ and $\mathcal{P}$ & If person $i$ reports $x_{iuv}=1$,\newline then there is a tie $\{i,k\}$ \newline in $\mathcal{M}$ for $\pi(uv)=k $\\ 

$ \mathbf{W}$ && Affiliation matrix of $\mathcal{M}$ &\\
$\mathbf{D}$ & $(D_r)_{r\in\mathcal{P} }$  & Colouring the nodes of $\mathcal{P}$ & if $\{u,v\} \in \mathcal{G}$, in criterion graph \newline $A_{uv}=1$ and for $\pi(uv)=r $ \newline $D_r=1$ \\

$\mathcal{Q}$ & & Graph on $\mathcal{P}$ & Ties between $u,v\in \mathcal{P}$ that share an element in the pre-image \\

$\mathbf{Q}$ & & Adjacency matrix of $\mathcal{Q}$ &\\

$\mathcal{C}$ && Network on $\mathcal{M}$,$\mathcal{Q}$, and $\mathcal{M}$ & One-mode representation of \newline multilevel network \\

$\mathbf{C}$ & & Blocked adjacency matrix of $\mathcal{C}$&\\

\hline
\end{tabular}
\end{center}
}}
\end{table}

\section{Goodness of fit}

\begin{table}
{\tiny{
\caption{\label{tab:MERGMGOF}Goodness-of-fit for Model 1}

\begin{center}
\begin{tabular}{lrrrr}
\hline

Statistics	&	Observed	&	Mean	&	StdDev	&	t-ratio	\\
EdgeA	&$	14	$&$	15.119	$&$	4.352	$&$	-0.257	$\\
Star2A	&$	27	$&$	32.059	$&$	21.424	$&$	-0.236	$\\
Star3A	&$	17	$&$	23.294	$&$	30.204	$&$	-0.208	$\\
Star4A	&$	6	$&$	13.501	$&$	33.548	$&$	-0.224	$\\
Star5A	&$	1	$&$	6.919	$&$	31.578	$&$	-0.187	$\\
TriangleA	&$	5	$&$	5.551	$&$	3.947	$&$	-0.14	$\\
Cycle4A	&$	5	$&$	5.299	$&$	7.802	$&$	-0.038	$\\
IsolatesA	&$	2	$&$	2.366	$&$	1.574	$&$	-0.233	$\\
IsolateEdgesA	&$	1	$&$	0.764	$&$	0.87	$&$	0.271	$\\
ASA	&$	19.875	$&$	23.0891	$&$	12.139	$&$	-0.265	$\\
ATA	&$	11.25	$&$	13.4626	$&$	7.878	$&$	-0.281	$\\
A2PA	&$	22.25	$&$	27.26	$&$	15.835	$&$	-0.316	$\\
AETA	&$	20	$&$	19.893	$&$	19.989	$&$	0.005	$\\
XEdge	&$	934	$&$	937.467	$&$	17.057	$&$	-0.203	$\\
XStar2A	&$	34071	$&$	33956.201	$&$	2422.932	$&$	0.047	$\\
XStar2B	&$	1361	$&$	1188.361	$&$	50.438	$&$	3.423	$\\
XStar3A	&$	897090	$&$	1013834.002	$&$	224473.748	$&$	-0.52	$\\
XStar3B	&$	1821	$&$	905.179	$&$	81.721	$&$	11.207	$\\
X3Path	&$	194167	$&$	169720.898	$&$	12961.535	$&$	1.886	$\\
X4Cycle	&$	11894	$&$	8735.213	$&$	1028.4	$&$	3.072	$\\
XECA	&$	1814656	$&$	1463003.616	$&$	316605.014	$&$	1.111	$\\
XECB	&$	89002	$&$	38742.874	$&$	5298.652	$&$	9.485	$\\
XASA	&$	1808.0156	$&$	1814.934	$&$	34.114	$&$	-0.203	$\\
XASB	&$	830.2759	$&$	835.8853	$&$	27.296	$&$	-0.206	$\\
XACA	&$	29078.7461	$&$	29889.6492	$&$	1990.761	$&$	-0.407	$\\
XACB	&$	195.2369	$&$	206.8379	$&$	1.395	$&$	-8.318	$\\
XAECA	&$	47575.4062	$&$	34940.852	$&$	4113.602	$&$	3.071	$\\
XAECB	&$	34082.3115	$&$	23951.0315	$&$	2965.459	$&$	3.416	$\\
Star2AX	&$	1742	$&$	1910.544	$&$	575.855	$&$	-0.293	$\\
StarAA1X	&$	1302.625	$&$	1469.0705	$&$	805.719	$&$	-0.207	$\\
StarAX1A	&$	3372.0156	$&$	3700.136	$&$	1118.421	$&$	-0.293	$\\
StarAXAA	&$	1864.0078	$&$	1875.41	$&$	40.86	$&$	-0.279	$\\
TriangleXAX	&$	185	$&$	205.084	$&$	66.178	$&$	-0.303	$\\
L3XAX	&$	51989	$&$	59569.904	$&$	19784.367	$&$	-0.383	$\\
ATXAX	&$	27.5154	$&$	30.077	$&$	8.665	$&$	-0.296	$\\
EXTA	&$	935	$&$	1065.106	$&$	785.434	$&$	-0.166	$\\
Star2BX	&$	24924	$&$	25017.347	$&$	460.361	$&$	-0.203	$\\
StarAB1X	&$	46413.8962	$&$	46587.6369	$&$	875.177	$&$	-0.199	$\\
StarAX1B	&$	25128.7515	$&$	24606.9828	$&$	801.756	$&$	0.651	$\\
StarAXAB	&$	15412.29	$&$	15423.0444	$&$	33.081	$&$	-0.325	$\\
TriangleXBX	&$	3740	$&$	3875.745	$&$	318.218	$&$	-0.427	$\\
L3XBX	&$	42069	$&$	42389.237	$&$	1577.281	$&$	-0.203	$\\
ATXBX	&$	2899.0625	$&$	3209.6011	$&$	241.658	$&$	-1.285	$\\
EXTB	&$	438305	$&$	440039.401	$&$	9601.51	$&$	-0.181	$\\
L3AXB	&$	45980	$&$	50570.331	$&$	15295.115	$&$	-0.3	$\\
C4AXB	&$	4900	$&$	5455.271	$&$	1852.387	$&$	-0.3	$\\
ASAXASB	&$	47716.5212	$&$	48056.7074	$&$	1231.73	$&$	-0.276	$\\
AC4AXB	&$	6372.4708	$&$	6902.4475	$&$	148.779	$&$	-3.562	$\\
stddev degreeA	&$	1.4573	$&$	1.3706	$&$	0.382	$&$	0.227	$\\
skew degreeA	&$	0.6445	$&$	0.2692	$&$	0.417	$&$	0.901	$\\
clusteringA	&$	0.5556	$&$	0.5144	$&$	0.156	$&$	0.264	$\\
stddev degreeX A	&$	70.2424	$&$	70.0838	$&$	2.472	$&$	0.064	$\\
skew degreeX A	&$	-1.103	$&$	-0.9459	$&$	0.044	$&$	-3.588	$\\
stddev degreeX B	&$	3.4014	$&$	3.238	$&$	0.057	$&$	2.875	$\\
skew degreeX B	&$	-0.9758	$&$	-1.1618	$&$	0.01	$&$	18.911	$\\
clusteringX	&$	0.245	$&$	0.2053	$&$	0.009	$&$	4.227	$\\
stddev degreeB	&$	20.9149	$&$	20.9149	$&$	0	$&$	-1	$\\
skew degreeB	&$	0.7521	$&$	0.7521	$&$	0	$&$	-1	$\\
clusteringB	&$	0.8341	$&$	0.8341	$&$	0	$&$	1	$\\
\hline
\end{tabular}
\end{center}
}}
\end{table}

\begin{table}
{\tiny{
\caption{\label{tab:MERGMGOFexpert}Goodness-of-fit for Model 2}
\begin{center}
\begin{tabular}{lrrrr}
\hline

Statistics	&	Observed	&	Mean	&	StdDev	&	t-ratio	\\
EdgeA	&	14	&	15.552	&	4.926	&	-0.315	\\
Star2A	&$	27	$&$	34.914	$&$	27.413	$&$	-0.289	$\\
Star3A	&$	17	$&$	31.167	$&$	50.992	$&$	-0.278	$\\
Star4A	&$	6	$&$	28.788	$&$	91.743	$&$	-0.248	$\\
Star5A	&$	1	$&$	28.722	$&$	151.577	$&$	-0.183	$\\
TriangleA	&$	5	$&$	5.594	$&$	4.37	$&$	-0.136	$\\
Cycle4A	&$	5	$&$	5.46	$&$	8.827	$&$	-0.052	$\\
IsolatesA	&$	2	$&$	2.183	$&$	1.45	$&$	-0.126	$\\
IsolateEdgesA	&$	1	$&$	0.719	$&$	0.868	$&$	0.324	$\\
ASA	&$	19.875	$&$	24.1388	$&$	14.163	$&$	-0.301	$\\
ASA2	&$	19.875	$&$	24.1388	$&$	14.163	$&$	-0.301	$\\
ATA	&$	11.25	$&$	13.6882	$&$	8.792	$&$	-0.277	$\\
A2PA	&$	22.25	$&$	29.9897	$&$	20.809	$&$	-0.372	$\\
AETA	&$	20	$&$	20.1698	$&$	22.405	$&$	-0.008	$\\
XEdge	&$	934	$&$	937.624	$&$	16.823	$&$	-0.215	$\\
XStar2A	&$	34071	$&$	33432.353	$&$	2609.927	$&$	0.245	$\\
XStar2B	&$	1361	$&$	1194.897	$&$	50.717	$&$	3.275	$\\
XStar3A	&$	897090	$&$	962121.188	$&$	234759.128	$&$	-0.277	$\\
XStar3B	&$	1821	$&$	929.591	$&$	86.868	$&$	10.262	$\\
X3Path	&$	194167	$&$	168319.628	$&$	14393.637	$&$	1.796	$\\
X4Cycle	&$	11894	$&$	8630.116	$&$	1170.167	$&$	2.789	$\\
XECA	&$	1814656	$&$	1410244.755	$&$	353314.367	$&$	1.145	$\\
XECB	&$	89002	$&$	39200.372	$&$	6229.365	$&$	7.995	$\\
IsolatesXA	&$	0	$&$	0	$&$	0	$&$	NaN	$\\
IsolatesXB	&$	0	$&$	0	$&$	0	$&$	NaN	$\\
XASA	&$	1808.0156	$&$	1815.248	$&$	33.646	$&$	-0.215	$\\
XASB	&$	830.2759	$&$	836.58	$&$	26.974	$&$	-0.234	$\\
XACA	&$	29078.7461	$&$	29424.9486	$&$	2115.042	$&$	-0.164	$\\
XACB	&$	195.2369	$&$	207.2175	$&$	1.278	$&$	-9.375	$\\
XAECA	&$	47575.4062	$&$	34520.464	$&$	4680.669	$&$	2.789	$\\
XAECB	&$	34082.3115	$&$	23788.9969	$&$	3385.866	$&$	3.04	$\\
Expert XEdgeA	&$	0	$&$	0	$&$	0	$&$	NaN	$\\
Expert XEdgeB	&$	359	$&$	361.437	$&$	10.688	$&$	-0.228	$\\
Expert X2StarA010	&$	0	$&$	0	$&$	0	$&$	NaN	$\\
Expert X2StarB010	&$	709	$&$	560.37	$&$	38.883	$&$	3.822	$\\
Expert X2StarA100	&$	0	$&$	0	$&$	0	$&$	NaN	$\\
Expert X2StarB100	&$	21247	$&$	21035.368	$&$	1660.846	$&$	0.127	$\\
Expert X2StarA101	&$	0	$&$	0	$&$	0	$&$	NaN	$\\
Expert X2StarB101	&$	5084	$&$	5115.007	$&$	463.06	$&$	-0.067	$\\
Expert X4CycleA1	&$	0	$&$	0	$&$	0	$&$	NaN	$\\
Expert X4CycleB1	&$	8943	$&$	6275.175	$&$	876.511	$&$	3.044	$\\
Expert X4CycleA2	&$	0	$&$	0	$&$	0	$&$	NaN	$\\
Expert X4CycleB2	&$	2997	$&$	1988.218	$&$	337.739	$&$	2.987	$\\
Expert XAlt4CycleA1	&$	0	$&$	0	$&$	0	$&$	NaN	$\\
Expert XAlt4CycleB1	&$	21543.7305	$&$	22368.1028	$&$	1605.882	$&$	-0.513	$\\
Expert XAlt4CycleA2	&$	0	$&$	0	$&$	0	$&$	NaN	$\\
Expert XAlt4CycleB2	&$	3948.6875	$&$	4220.5817	$&$	329.981	$&$	-0.824	$\\
Expert XEdgeAB	&$	0	$&$	0	$&$	0	$&$	NaN	$\\
Star2AX	&$	1742	$&$	1999.047	$&$	826.083	$&$	-0.311	$\\
StarAA1X	&$	1302.625	$&$	1666.4692	$&$	1244.166	$&$	-0.292	$\\
StarAX1A	&$	3372.0156	$&$	3873.678	$&$	1614.366	$&$	-0.311	$\\
StarAXAA	&$	1864.0078	$&$	1877.456	$&$	42.567	$&$	-0.316	$\\
TriangleXAX	&$	185	$&$	217.22	$&$	104.594	$&$	-0.308	$\\
L3XAX	&$	51989	$&$	63034.626	$&$	31728.236	$&$	-0.348	$\\
ATXAX	&$	27.5154	$&$	30.9576	$&$	9.829	$&$	-0.35	$\\
EXTA	&$	935	$&$	1144.756	$&$	1065.915	$&$	-0.197	$\\
Star2BX	&$	24924	$&$	25079.4	$&$	426.565	$&$	-0.364	$\\
StarAB1X	&$	46413.8962	$&$	46709.5707	$&$	810.195	$&$	-0.365	$\\
StarAX1B	&$	25128.7515	$&$	24736.3993	$&$	751.933	$&$	0.522	$\\
StarAXAB	&$	15412.29	$&$	15423.3979	$&$	32.453	$&$	-0.342	$\\
TriangleXBX	&$	3740	$&$	3814.606	$&$	353.514	$&$	-0.211	$\\
L3XBX	&$	42069	$&$	42598.521	$&$	1462.718	$&$	-0.362	$\\
ATXBX	&$	2899.0625	$&$	3153.7942	$&$	265.147	$&$	-0.961	$\\
EXTB	&$	438305	$&$	441502.275	$&$	8880.053	$&$	-0.36	$\\
Expert Star2BX	&$	45980	$&$	53366.561	$&$	23225.159	$&$	-0.318	$\\
L3AXB	&$	4900	$&$	5874.284	$&$	3164.443	$&$	-0.308	$\\
C4AXB	&$	47716.5212	$&$	48376.0399	$&$	1509.633	$&$	-0.437	$\\
stddev degreeA	&$	1.4573	$&$	1.3937	$&$	0.445	$&$	0.143	$\\
skew degreeA	&$	0.6445	$&$	0.3555	$&$	0.538	$&$	0.537	$\\
clusteringA	&$	0.5556	$&$	0.4813	$&$	0.15	$&$	0.494	$\\
stddev degreeX A	&$	70.2424	$&$	69.5409	$&$	2.663	$&$	0.263	$\\
skew degreeX A	&$	-1.103	$&$	-0.9516	$&$	0.04	$&$	-3.787	$\\
stddev degreeX B	&$	3.4014	$&$	3.2445	$&$	0.057	$&$	2.761	$\\
skew degreeX B	&$	-0.9758	$&$	-1.1588	$&$	0.011	$&$	17.01	$\\
clusteringX	&$	0.245	$&$	0.2043	$&$	0.011	$&$	3.77	$\\
stddev degreeB	&$	20.9149	$&$	20.9149	$&$	0	$&$	-1	$\\
skew degreeB	&$	0.7521	$&$	0.7521	$&$	0	$&$	-1	$\\
clusteringB	&$	0.8341	$&$	0.8341	$&$	0	$&$	1	$\\

\hline
\end{tabular}
\end{center}
}}
\end{table}

\end{document}